\documentclass[%
 aip,
 floatfix,
 amsmath,amssymb,
 reprint,%
]{revtex4-1}

\usepackage{graphicx}
\usepackage{dcolumn}
\usepackage{bm}

\usepackage[utf8]{inputenc}
\usepackage[T1]{fontenc}
\usepackage{mathptmx}
\usepackage{siunitx}
\usepackage{subfig}
\usepackage{color}

\begin{document}

\preprint{AIP/123-QED}

\title[Radiative models for simulation of lightning strikes]{Equation
  of state driven radiative models for simulation of lightning strikes}

\author{M. Apsley}
\email{ma616@cam.ac.uk}
\author{S. T. Millmore}%
\author{N. Nikiforakis}
\affiliation{ 
Cavendish Laboratory, Department of Physics, University of Cambridge, Cambridge, United Kingdom
}%


\date{\today}

\begin{abstract}
  This work is concerned with the numerical simulation of plasma arc
  interaction with aerospace substrates under conditions akin to
  lightning strike and in particular with the accurate calculation of
  radiative heat losses. These are important because they have a
  direct effect on the calculation of thermal and pressure loads on
  the substrates, which can lead to material damage under certain
  conditions.  Direct numerical solution of the radiation transport
  equation (RTE) in mesoscale simulations is not viable due to its
  computational cost, so for practical applications reduced models are
  usually employed. To this end, four approximations for solving the
  RTE are considered in this work, ranging from a simple local
  thermodynamical behaviour consideration, to a more complex spectral
  absorption dependent on the arc geometry.  Their performance is
  initially tested on a one-dimensional cylindrical arc, before
  implementing them in a multi-dimensional magnetohydrodynamics
  code. Results indicate that inclusion of spectral absorption is
  necessary in order to obtain consistent results. However, the
  approaches accounting for the arc geometry require repeated solution
  of the computationally intensive Helmholtz equations, making them
  prohibitive for multi-dimensional simulations.  As an alternative, a
  method using the net emission coefficient is employed, which
  provides a balance between computational efficiency and accuracy, as
  shown by comparisons against experimental measurements for a plasma
  arc attaching to an aluminium substrate.
\end{abstract}

\maketitle

\section{Introduction}

Lightning strikes last for $\mathcal{O}(100)$ \si{\milli \second},
producing currents of $\mathcal{O}(10)$ \si{\kilo \ampere} in pulses
lasting for $\mathcal{O}(100)$ \si{\mu \second}
\cite{martins2016etude}. The resulting ionisation generates magnetic
fields, coupled to the lightning arc through Joule heating and the
Lorentz force, and this injects heat into the plasma, increasing the
temperature by \SI{30}{\kilo \kelvin} over the first few microseconds
\cite{chemartin2012direct}.  Over these short time scales, the heat is
dissipated through a combination of radiation and convection and, in
the case of lightning striking an aircraft, energy is also dissipated
by Joule heating within the aircraft material as the electric fields
circulate through it.

In the case of lightning attachment to aircraft, there is particular
concern for modern designs utilising lightweight composite materials.
These materials have much lower thermal and electrical conductivity
than traditional materials, such as aluminium, and can undergo
delamination and fibre breakage as a result~\cite{zhang2019lightning,
  chemartin2012direct}.  Numerical models offer a predictive tool
which can capture Joule heating and subsequent damage to aircraft
substrates, but to accurately capture the behaviour over the early
stages of arc attachment, any model implemented relies on accurately
modelling the current transferred to the substrate from the arc.
Current density is greatest at the centre of the arc, hence any
numerical model must be able to predict the plasma dynamics in this
region.  Due to the high temperatures, radiative heat losses have a
significant contribution to the underlying evolution, and thus an
accurate radiative model is essential for modelling Joule heating
effects in simulating lightning attachment to aircraft substrates.  At
the centre of the arc, overpressure significantly increases radiative
emission which in turn limits further growth of pressure, and
temperature, hence an accurate model is essential to avoid
over-prediction of these quantities.  However, in addition to
pressure, radiative emission is also dependent on the composition,
temperature, density and geometry of the plasma, and subsequent
absorption in the peripheral regions of the arc is also known to be
very important to the rate of arc expansion.

A full description of the radiative losses within a plasma arc
requires solving for energy balance at every point in the arc,
spanning the full range of wavelengths in the electromagnetic
radiation.  These radiative effects are described in full by the
radiative transfer equation (RTE) and this has been solved fully for
stationary cylindrically symmetric lightning arcs by Lowke
\cite{Lowke1969Relaxation}, Lowke and Capriotti
\cite{lowke1969calculation} and Hermann and Schade
\cite{hermann1972radiative}. However for dynamic simulations, the
exact solution of the RTE is prohibitively slow and numerical
simplifications have been sought
\cite{Chauveau2003RadiativeTransfer}.  A key challenge here is to
balance the accuracy of the method with computational efficiency.
Solutions of the RTE, approximate or exact, require knowledge of the
spectral properties of the plasma; this information is encapsulated
within the monochromatic absorption coefficient and is a measure of
the amount of radiation absorbed by the plasma at any given frequency
\cite{cressault2015basic}.

Approximate solutions of the RTE will be collectively referred to in
this paper as spectral methods. The simplest approximate solution is
the net emission coefficient (NEC) introduced by Lowke
\cite{Lowke1974Predictions}, which considers only the radiation
emitted by the plasma and, in its simplest form, all self-absorption
is neglected.  Despite its simplicity it has been found to produce
good predictions for the volumetric radiative power in the hottest
parts of the plasma, but is less accurate near boundaries or in
regions of high pressure \cite{raynal1995radiative,eby1998modelling}.
It is implemented by e.g. Chemartin et al. \cite{chemartin2009three},
Teulet et al. \cite{teulet2017energy} and Fusheng et
al. \cite{fusheng2018evolution}.  It has also been applied to
modelling lightning streamers by e.g. Da Silva et
al. \cite{da2019plasma}.

More accurate than the NEC are the P$_N$ approximations which were
introduced by Lacombe et al.~\cite{lacombe2008role} and used by
e.g. Aleksandrov \cite{aleksandrov2000effect} and Tholin et
al.~\cite{tholin2015numerical}. They are spherical harmonic
expansions which satisfy the RTE to $\mathcal{O}(N)$ accuracy.  In
between these two approaches, the SP$_N$ approximations simplify the
complexity of the P$_N$ equations significantly, allowing higher order
of accuracy solutions to be calculated for the computational effort.
As the radiative emission is wavelength dependent, to reduce the
number of calculations, solutions are found for a number of averaged
bands in the emission spectrum e.g. $11$ in Peyrou et
al.~\cite{peyrou2012radiative}, Bartlova et 
al. ~\cite{bartlova2017modelling} and Tholin et
al.~\cite{tholin2015numerical}, $5$ in Eby et
al.~\cite{Eby1998P1Approximation}, and $4$ in Aleksandrov et
al.~\cite{aleksandrov2000effect}.  Nordborg and Iordanidis
\cite{nordborg2008self} compare the accuracy of different methods of
averaging the absorption coefficient in an SF$_6$ plasma and find that
averaged models are sufficiently accurate so long as averages are
calculated with a weighting based on their magnitude.

Higher accuracy approximate solutions of the RTE do exist, including
the method of partial characteristics introduced by Sevast'yaneko
\cite{sevast1979radiation,sevast1980radiation} which tabulates the
radiative exchange between volume elements in the plasma in advance by
assuming a functional form between each pair of elements. This
requires significantly more precalculation than any of the previous
methods.  Alternatively Monte-Carlo methods, such as those described
by Gogel et al.~ \cite{gogel1994radiation} can be used but these are
computationally expensive and impractical for coupling with a full
magnetohydrodynamic description \cite{peyrou2012radiative}.

Within this work, four spectral methods are considered, with their
formulation for use in one-dimensional lightning arc simulations
evaluated.  The NEC, P$_1$ and SP$_3$ are considered, along with an
alternative NEC approach considered by Villa et
al.~\cite{villa2011multiscale}.  This latter approach treats the
plasma as a grey body of constant emissivity and solves a simplified
RTE using the temperature, density and at specific heat capacity of
the plasma at a given point.  The radiative emission is not dependent
on the properties of the neighbouring plasma and hence this method can
be thought of as an alternative NEC method \cite{martins2016etude}.
These methods are implemented within a validated magnetohydrodynamics
(MHD) code developed by Millmore and
Nikiforakis~\cite{millmore2019multi} using an accurate equation of
state (EoS) developed by Tr{\"a}uble et
al.~\cite{trauble2021improved}.  The theoretical background of these
approaches is detailed in section~\ref{Sec:Physics of a lightning
  strike}, and the results for these one-dimensional implementations
given in section~\ref{Sec:NumericalResults}.  Based on these results,
the implementation of these techniques for efficient multi-dimensional
cases is considered.  In particular, the NEC, and an improved reduced
model, considering an optically dense plasma, are implemented such
that the approximate solution to the RTE can be considered a function
of the EoS.  This allows these approaches to be incorporated in the
tabulated data for the equation of state introduced by Tr{\"a}uble et
al.  Results for these multi-dimensional effects are shown, and
validated in section~\ref{Sec:GroupCode}, considering a test case of
arc attachment to an aluminium substrate.  Finally, conclusions and
future work are described in section~\ref{Sec:Conclusions}.

\section{Mathematical models}
\label{Sec:Physics of a lightning strike}  

\subsection{Plasma evolution equations}
\label{sec:plasma-evol-equat}

The evolution of a lightning strike is governed by coupling the Euler
equations describing the conservation of density $\rho$, momentum
$\rho \textbf{u}$, and energy $E$, with the Maxwell equations
governing the electromagnetic interaction of the magnetic field
$\textbf{B}$, and the current density
$\textbf{j}$~\cite{millmore2019multi,michael2019multi}. The
electromagnetic coupling occurs over much faster timescales than the
fluid evolution and hence the quasi-static approximation to Maxwell's
equations can be used
\cite{chemartin2009three,tanaka2005hydrodynamic}.  In this limit, the
magnetic field is given by
\begin{equation}
  \textbf{B} = \nabla \times \textbf{A}
\end{equation} 
where $\textbf{A}$ is the vector magnetic potential.  In a material of
electrical conductivity $\sigma$, the equations governing the time
evolution are
\begin{equation}
 \frac{\partial \rho}{\partial t} + \nabla \cdot (\rho \textbf{u}) = 0,
 \label{Eq:MassEvolution}
\end{equation}
\begin{equation}
 \frac{\partial \rho \textbf{u}}{\partial t} + \nabla \cdot(\rho \textbf{u} \textbf{u}^T + p \mathcal{I}) = \textbf{j} \times \textbf{B},
\end{equation}
\begin{equation}
 \frac{\partial E}{\partial t} + \nabla \cdot (\textbf{u}(E + p)) = \frac{\textbf{j}^2}{\sigma} - S_R + \textbf{u} \cdot (\textbf{j} \times \textbf{B}),
\end{equation}
\begin{equation}
  -\nabla^2 \textbf{A} = \mu_0 \textbf{j},
  \label{Eq:CurrentPoisson}
\end{equation}
where $\mu_0$ is the permeability of free space.  This approach
assumes local thermodynamic equilibrium (LTE), assuming that the
ionised material can be described with a single local temperature and
pressure, which has been demonstrated to be able to accurately predict
the behaviour of a lightning
plasma~\cite{Gleizes2005ThermalPlasmaModelling,haidar1999non}.

To close this system of equations, boundary conditions for current
density, a radiative model supplying the source term $S_R$, and an EoS
describing the interdependence of internal energy $e$, pressure $p$,
and temperature $T$ are required.  In this work, boundary conditions
for current density come from a prescribed current profile at the top
of the computational domain, provided by standard wave forms for
aerospace testing~\cite{arp5412a2005aircraft}, and a 19-species EoS
tabulated and validated by~\cite{trauble2021improved} is used.  The
choice of radiative source term is the focus of this work, but it is
worth noting that in addition to closing the system of
equations~(\ref{Eq:MassEvolution})--(\ref{Eq:CurrentPoisson}), it can
further be used to calculate the frequency-dependent monochromatic
absorption coefficient of light, determined by the interaction of
photons with the plasma, which is required by the NEC, P$_1$, and
SP$_N$ models to calculate the radiation source term.

A full description of the numerical techniques for solving
equations~(\ref{Eq:MassEvolution}) to (\ref{Eq:CurrentPoisson}) is
given in Millmore and Nikiforakis~\cite{millmore2019multi}.  This is a
four-step process, solving in the first step the Poisson's equations
for the electric potential and magnetic vector potential, and in the
second step, solving the evolution equations without forcing terms. In
the third step the energy in the plasma is updated to account for
radiation and Joule heating, and the final step couples the Lorentz
force to the fluid evolution.  The general approach to including the
radiative source term in the third step is not dependent on the method
used to obtain this term, in all cases an ordinary differential
equation (ODE) is used to update the energy,
\begin{equation}
  \frac{\partial E}{\partial t} = \frac{\textbf{j}^2}{\sigma} - S_R.
  \label{Eq:EnergyUpdate}
\end{equation}
For all methods considered, except the grey body approach, the
radiative term, $S_R$, is computed directly, and then standard ODE
solvers are used to update the energy.  For the grey body model, the
solution to the ODE is coupled to the approximate solution to the RTE,
this method is described in section \ref{Sec:GreyBodyModel}.

\subsection{Grey body model}
\label{Sec:GreyBodyModel}

The grey body model is of simplified NEC model used by Villa et
al.~\cite{villa2011multiscale}. Although this approach simplifies the
underlying model for solving the RTE, by neglecting spectral and
geometric dependencies, it does add complexity in the techniques for
solving equation~(\ref{Eq:EnergyUpdate}); this process can be broken
down into three steps.  In the first, the plasma is assumed to absorb
energy from its surroundings which are at an ambient temperature,
$T_0$.  For a grey body of emissivity $\kappa_{emiss}$ the energy
absorbed is given by $\kappa_{emiss} \sigma_{SB} T_0^4$ where
$\sigma_{SB}$ is the Stefan-Boltzmann constant and Villa et al. use
$\kappa_{emiss} = 60.0$.  This absorption is treated as a source term
alongside the Joule heating, and the energy is updated using the
first-order explicit step,
\begin{equation}
  E^{n+1/2} = E^{n} + \Delta t \left( \frac{\textbf{j}^2}{\sigma} + \kappa_{emiss} \sigma_{SB} T_0^4 \right)
\end{equation}
where $\Delta t$ is the discretised time step of the simulation.

In the second step an intermediate temperature, $T^{n+1/2}$,
corresponding to this updated energy, is calculated, and from this
temperature, and the EoS, mass fractions of the chemical species are
obtained.  From these the species-averaged heat capacity and heat of
formation heats must be determined; the specific heat capacity at
constant volume, $c_v^i$, for a diatomic species of molar mass $M_k$
depends on the specific gas constant $R_k = \frac{\hat{R}}{M_k}$ and
the temperature $T$,
\begin{equation}
  c_v^i (T) = \left( \frac{5}{2} + \left( \frac{T_M}{T} \right)^2 \frac{\exp(T_M/T)}{(\exp(T_M/T) - 1)^2} \right) R_k,
\end{equation}
where $T_M$ is a characteristic temperature for the atomic species
involved. For a monatomic species, $c_v^i$ is given by:
\begin{equation}
  c_v^i (T) = \frac{3}{2} R_k.
\end{equation}
The specific heats of formation for each species,
$\Delta h_f^{0 , i}$, are taken from thermochemical
tables~\cite{tables1985parts,ruscic2017active,WebElements}.  The
species-averaged values are then calculated from these quantities, and
the mass fractions, $x_i$, as provided by the EoS, according to
\begin{equation}
 c_v = \sum_{\text{species}} c_v^i (T^{n+1/2}) x_i,
\end{equation}
and
\begin{equation}
\Delta h_f^0 = \sum_{\text{species}} (\Delta h_f^0)^i x_i.
\end{equation}
From these, the specific heat capacity at constant pressure, $c_p$, as well as
the corresponding temperature associated with the heat increase having
taken place at constant pressure, $T_b$, can be calculated:
\begin{equation}
  c_p = \alpha c_v
\end{equation}
and
\begin{equation}
  T_{b} = \alpha T^{n+1/2},
\end{equation}
where
\begin{equation}
  \alpha = \frac{(e - \Delta h^f_0)}{c_v T^{n+1/2}}.
\end{equation}

The final step calculates the temperature of the plasma after
radiative emission.  Again the plasma is modelled as a grey body, and
emission is described by
\begin{equation}
  \rho c_p \frac{dT}{dt} = - \sigma_{SB} \kappa_{emiss} T^4.
\end{equation}
This simplified RTE is solved analytically by rearranging and
integrating from time $t$ to $t + \Delta t$, to obtain an updated
temperature, $T_{nb}$. This produces an explicit expression for
$T_{nb}$,
\begin{equation}
  T_{nb} = \left( T_b^{-3} + \frac{3 \sigma_{SB}\kappa_{emiss} \Delta t}{\rho c_p} \right)^{-1/3},
\end{equation}
and from this, the total energy at temperature $T_{nb}$
is found by adding the enthalpic, thermal and kinetic contributions,
\begin{equation}
  E^{n+1} = \rho^{n+1} (\Delta h_0^f + c_p T_{nb} + \frac{1}{2} (u^{n+1})^2)
\end{equation}

\subsection{Net emission coefficient}
\label{Sec:NEC}

The NEC improves upon the grey body approach by considering the
spectral dependencies of the radiative emission.  The radiation
intensity at a given frequency per unit solid angle, $I_{\nu}$, is
calculated by solving the RTE which, for spherical symmetry in a
non-scattering plasma with unit refractive index, is:
\begin{equation}
  \frac{\partial I_{\nu}}{\partial r} = \kappa'_\nu B_{\nu} - \kappa'_{\nu}I_{\nu}.
  \label{Eq:RTEradial}
\end{equation}
where $\kappa'_{\nu}(T,p)$ is the modified absorption coefficient
taking into account stimulated emission,~\cite{RaynalGleizes1995}
\begin{equation}
  \kappa'_{\nu} = \kappa_{\nu} \left[ 1 - \exp \left(- \frac{h\nu}{k_{B} T} \right) \right],
\end{equation}
with $h$ and $k_B$ being the Planck and Boltzmann constants
respectively, and the spectral radiation intensity $B_{\nu}$ is given by Planck's law for black body
radiation,
\begin{equation}
  B_{\nu} = \frac{2h\nu^3}{c^2 \left[ \exp\left( \frac{h\nu}{k_{B}T} \right) - 1 \right]},
\end{equation}
where $c$ is the speed of light. Under the assumption of LTE, Kirchoff's law holds, hence the emission
coefficient at frequency $\nu$, $\varepsilon_{\nu}$, satisfies
\begin{equation}
  \frac{\varepsilon_{\nu}}{\kappa'_{\nu}} = B_{\nu}.
\end{equation}
The NEC, $\varepsilon_N$, is then given by integrating over all
frequencies for a plasma contained within an isothermal sphere of
radius $R_P$,
\begin{equation}
  \varepsilon_N(T,p,R_p) = \int_0^{\infty} B_{\nu} \kappa'_{\nu} \exp(-\kappa'_{\nu}R_P) d\nu,
  \label{Eq:NECequation}
\end{equation}
where $R_p$ is the decay length of the plasma. An assumption of
$R_P = 0$ is equivalent to assuming an optically thin plasma for which
about $90$\% of the radiation is absorbed.  Within this paper,
$R_p = 0$ is referred to as the NEC model, and non-zero values of
$R_p$ are reduced NEC models. In order to obtain the radiative source
term from this value, the NEC is related to the divergence of
radiative flux $\textbf{q}$ and hence $S_R$ by
\cite{Lowke1974Predictions,Lowke1969Relaxation}
\begin{equation}
  S_R = \nabla \cdot \textbf{q} = 4\pi \varepsilon_N.
  \label{Eq:NECsource}
\end{equation}
Computationally, computing the integral in
equation~(\ref{Eq:NECequation}) requires a discretisation of the
frequency, $\nu$.  Following Nordborg and
Indianis~\cite{nordborg2008self}, the absorption coefficient is
computed using band averaged values of $\kappa_{\nu}$ and
$B_{\nu} \kappa_{\nu}$, denoted $\bar{\kappa_i}$ and
$\bar{(B\kappa )_i}$ respectively, and a Planck-average is used to
compute these values, for example
\begin{equation}
  \bar{\kappa_i} = \frac{1}{\int_{\Delta \nu_i} B_{\nu} d\nu} \int_{\Delta \nu} B_{\nu} \kappa_{\nu} d\nu,
  \label{eq:1}
\end{equation}
where the frequency range is given by $\Delta \nu$.  It is sufficient
for five band-averaged values to be used to compute $\varepsilon_N$,
hence the radiative source term is given by
\begin{equation}
  S_R = 4 \pi \sum_{i=1}^5 \bar{(B \kappa )_i} \exp (-\bar{\kappa_i} R_p).
  \label{eq:NECSourceNum}
\end{equation}
Further details for computing these quantities are given in
appendix~\ref{Sec:AbsorptionCoefficient}.

For the NEC and reduced NEC models, further computational gains are
available since the NEC itself only depends on local pressure and
temperature.  These quantities can then be computed as part of the
tabulated EoS and interpolated values used to compute radiative losses.

\subsection{P$_1$ approximation}

The P$_N$ approximations improve upon the NEC by beginning to account
for geometric effects which lead to self-absorption of radiation by
the plasma. The derivation of the P$_N$ approximations starts from the
directional RTE, neglecting scattering in the direction
$\hat{\textbf{s}}$ \cite{Eby1998P1Approximation},
\begin{equation}
  \hat{\textbf{s}} \cdot \nabla I_{\nu} = \kappa_{\nu} (B_{\nu} - I_{\nu}).
  \label{Eq:DirectionalRTE}
\end{equation}
The radiation intensity is decomposed into direction-dependent
coefficients of the spherical harmonics $Y_l^m(\hat{\textbf{s}})$ and
spatial-dependent coefficients $P_l^m (\textbf{r})$:
\begin{equation}
  I_{\nu}(\mathbf{r,s}) = \sum_{l = 0}^{\infty} \sum_{m=-l}^{l} P_l^m (\textbf{r}) Y_l^m(\hat{\textbf{s}}).
\end{equation}
Due to the complexities of evaluating this expression for higher-order
approximations, in this work only the $P_1$ approximation is used, for
which only the first term in the expansion is kept.  Eby et al
\cite{Eby1998P1Approximation} show that this is of the form
\begin{equation}
  I_{\nu}(\textbf{r},\textbf{s}) = \frac{1}{4\pi} (G_{\nu} + 3 \textbf{q}_{\nu} \cdot \hat{\textbf{s}}).
  \label{Eq:P1Intensity}
\end{equation}
where $G_{\nu}$ is the incident radiation defined by integrating over solid angle $\Omega$,
\begin{equation}
  G_{\nu}  = \int_{4\pi} I_{\nu}(\mathbf{r,\hat{s}}) d\Omega,
\end{equation}
and $\textbf{q}_{\nu}$ is the radiative flux,
\begin{equation}
  \textbf{q}_{\nu}  = \int_{4\pi} I_{\nu}(\textbf{r},\hat{\textbf{s}}) \hat{\textbf{s}} d\Omega.
\end{equation}
Substitution of equation~\eqref{Eq:P1Intensity} into
equation~\eqref{Eq:DirectionalRTE} generates the governing equations,
\begin{equation}
  \nabla G_{\nu} = -3\kappa_{\nu}\textbf{q}_{\nu}
  \label{Equ:P1Approximation1st}
\end{equation}
and
\begin{equation}
  \nabla \cdot \textbf{q}_{\nu} = \kappa_{\nu} (4\pi B_{\nu} - G_{\nu}).
  \label{Equ:P1Approximation2nd}
\end{equation}
These can be combined into a single Helmholtz-type equation for $G_{\nu}$,
\begin{equation}
  \nabla \cdot \left( \frac{1}{\kappa_{\nu}} \nabla G_{\nu} \right) = 3 \kappa_{\nu}(G_{\nu} - 4 \pi B_{\nu}).
  \label{Eq:P1HelmholtzEquation}
\end{equation}
As with the NEC approach, five frequency bands are used for solving
this equation, and the volumetric radiation source term is given by
\begin{equation}
  S_R = \nabla \cdot \bar{\textbf{q}} = \sum_{i=1}^5 4 \pi \bar{(\kappa B)_i} - \bar{\kappa_i} \bar{G_i}
  \label{Eq:P1Source}
\end{equation}
Unlike the NEC approach, the geometric dependencies of
equation~(\ref{Eq:P1Source}) mean that the radiative losses cannot be
included within the EoS data.  In this work, the solution to this
equation uses a five-point stencil, to produce a block tridiagonal
system, and is achieved through the ADI method
\cite{wachspress2008trail}.

\subsection{SP$_3$ Approximation}
\label{Sec:SP3Approximation}

To obtain higher order P$_N$ approximations the complexity and number
of equations increases.  The SP$_N$ approximations are a way of
simplifying these equations and reducing them in number for the case
of an optically thick plasma with length scale $x_{\mathrm{ref}}$ and
characteristic opacity $\kappa_{\mathrm{ref}}$ such that
\begin{equation}
  \varepsilon = \frac{1}{\kappa_{ref} x_{ref}}
\end{equation}
satisfies $0 < \varepsilon \ll 1$.  Assuming that thermal diffusion
occurs with thermal conductivity $k_T$, the energy balance equation
can be written
\begin{equation}
  \rho c_p \frac{\partial T}{\partial t} = \nabla \cdot (k_T \nabla T) - \int_{\nu_1}^{\infty} \int_{\mathcal{S}^2} \kappa_{\nu} (B_{\nu} - I_{\nu}) d\Omega d\nu,
\end{equation}
and the directional RTE (\ref{Eq:DirectionalRTE}) can be rescaled
\cite{Larssen2002SPNApproximation}, giving
\begin{equation}
  \varepsilon^2 \frac{\partial T}{\partial t} = \varepsilon^2 \nabla \cdot (k_T \nabla T) - \int_{\nu_1}^{\infty} \int_{\mathcal{S}^2} \kappa_{\nu} (B_{\nu} - I_{\nu}) d\Omega d\nu
  \label{Eq:EnergyBalance}
\end{equation}
and
\begin{equation}
  \forall \nu > \nu_1, \quad \Omega \in \mathcal{S}^2\text{:    } \varepsilon \Omega \cdot \nabla I = \kappa_{\nu} (B_{\nu} - I_{\nu}).
  \label{Eq:EqnOfTransfer}
\end{equation}
This latter equation can be rearranged to give
\begin{equation}
  I_{\nu} =  \left(1 - \frac{\varepsilon}{\kappa_{\nu}}\Omega \cdot \nabla \right)^{-1} B_{\nu}
\end{equation}
and then expanded in powers of
$\frac{\varepsilon}{\kappa_{\nu}} \Omega \cdot \nabla$.  By introducing the integral of the radiation intensity at given freqency over solid angles,
\begin{equation}
  \psi = \int_{\mathcal{S}^2} I_{\nu} d \Omega,
\end{equation}
retaining terms up to $\mathcal{O}(\varepsilon^8)$, substituting into
equation~\eqref{Eq:EnergyBalance}, and simplifying, two expressions
are obtained,\cite{Larssen2002SPNApproximation}
\begin{equation}
  -\nabla \cdot \left( \frac{\mu_1}{\kappa_{\nu}} \nabla \psi_{1,\nu} \right) + \kappa_{\nu} \psi_{1,\nu} = (4 \pi B_{\nu})\kappa_{\nu}
  \label{eq:SP3Eq1}
\end{equation}
\begin{equation}
  -\nabla \cdot \left( \frac{\mu_2}{\kappa_{\nu}} \nabla \psi_{2,\nu} \right) + \kappa_{\nu} \psi_{2,\nu} = (4 \pi B_{\nu})\kappa_{\nu}
  \label{eq:SP3Eq2}
\end{equation}
where $7\mu_{1,2} = 3 \mp \sqrt{6/5}$.  In this model the following
boundary conditions were assumed \cite{Eby1998P1Approximation}:
\begin{itemize}
  \item A symmetry condition: $\partial_r G_{\nu}|_{r = 0} = 0$,
  \item No flux at the air boundaries: $\textbf{n} \cdot \nabla G_{\nu} = 0$
  \item No radiation transfer to the solid surfaces: $\textbf{n} \cdot \nabla G_{\nu} = 0$
\end{itemize}
These boundary conditions decouple equations \eqref{eq:SP3Eq1} and
\eqref{eq:SP3Eq2}, which can then be treated as Helmholtz equations,
and are again discretised using the ADI method with 5 bands used for
the frequency averaging of these equations. For the SP$_3$
approximation, this then requires 10 Helmholtz equations to be solved
each time step.  The radiative power for the SP$_3$ approximation is
given by
\begin{equation}
  S_R = \int_0^{\infty} \nabla \cdot \left( \frac{1}{\kappa_{\nu}} \nabla (a_1\psi_{1,\nu} + a_2 \psi_{2,\nu}) \right)d\nu
  \label{Eq:SPNsource}
\end{equation}
where $30 a_{1,2} = (5 \mp 3\sqrt{5/6})$, \cite{peyrou2012radiative}
and again, this quantity is approximated by a sum over the
band-averaged values,
\begin{equation}
  S_R = \sum_{i = 1}^5 \frac{a_1}{\mu_1}(\bar{\kappa \psi}_{1,i} - 4 \pi \bar{\kappa B}_i )  + \frac{a_2}{\mu_2}(\bar{\kappa \psi}_{2,i} - 4 \pi \bar{\kappa B}_i ) 
\end{equation}
Again, the geometric dependency of the SP$_3$ method means that these
values cannot be tabulated as part of the EoS data.

\section{Numerical Results}
\label{Sec:NumericalResults}

In order to compare the different radiative models, an initial
assessment was made using a one-dimensional, cylindrical plasma arc.
This follows experimental and numerical studies of Villa et
al.~\cite{villa2011multiscale}, and was used to validate the
underlying numerical framework by Millmore and
Nikiforakis~\cite{millmore2019multi}.  Experimentally, a plasma arc
attaches to a highly conductive metal substrate within an set up large
enough such that the arc can be treated as cylindrical.  The arc is
driven by an input current from an electrode with a profile given by
\begin{equation}
  I = I_0 \exp(-\alpha t) sin (\beta t).
\end{equation}
The simulation is initialised with a high-temperature pre-heated arc
region to approximate the initial breakdown of air generating this
structure.  This approach avoids the need to incorporate the complex
physics of the breakdown itself, and has been shown by Tholin et
al.~\cite{tholin2015numerical} not to affect the overall evolution of
the arc.  The initial conditions for this test case are given in table
\ref{Tab:ICfor1Dtest}, and represent a preheated arc of radius
\SI{2}{\centi \metre} with a Gaussian temperature profile and maximum
temperature \SI{8680}{\kelvin}. The simulation was implemented in the
domain $ x \in [0, 25]$ \si{cm} and, following the
method of Villa et al, a current density profile of the form
$\textbf{J} = J(\textbf{r}) \textbf{e}_z$ was used. The remaining
initial conditions are given in table~\ref{Tab:ICfor1Dtest}.

\begin{table}[!hbtp]
\begin{center}
\begin{tabular}{| c | c | c |}
  \hline
   & $r < 0.02$ & $r > 0.02$ \\ \hline
  $\rho$ (\si{kg m^{-3}}) & $0.9$ & $1.225$  \\ \hline
  $u_r$ (\si{m s^{-1}}) & $100$ & $0$ \\ \hline
  $E$ (\si{J}) & $20 \times 335014 \exp \left(- \frac{x^2}{0.5^2} \right) $ & $335014$ \\ \hline
\end{tabular}
\caption{Initial conditions for a cylindrical plasma arc, reproducing
  the experiment of Villa et al.~\cite{villa2011multiscale}.}
\label{Tab:ICfor1Dtest} 
\end{center}
\end{table}

\begin{figure}[!hptb]
\centering
\subfloat[\SI{20}{\micro \second} ]{\includegraphics[scale=.4]{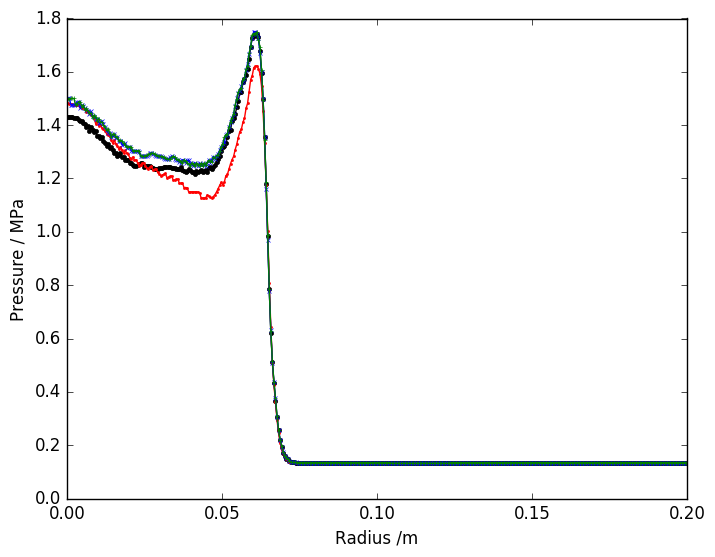} }\\
\subfloat[\SI{60}{\micro \second} ]{\includegraphics[scale=.4]{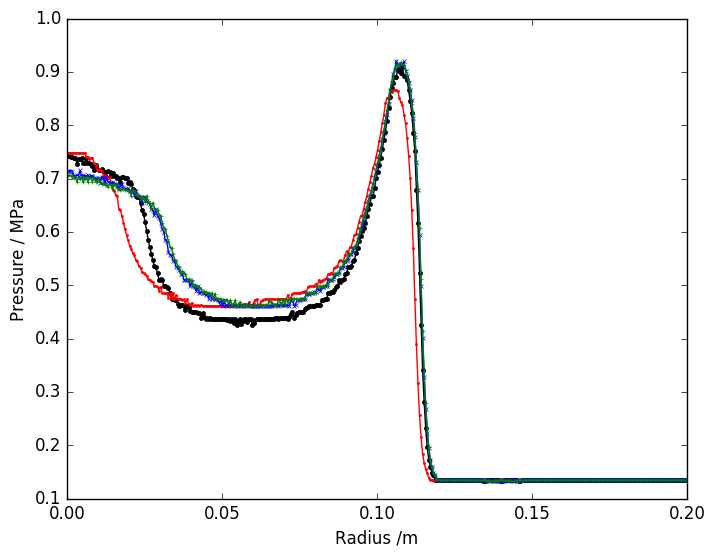} }\\
\subfloat[\SI{100}{\micro \second}]{\includegraphics[scale=.4]{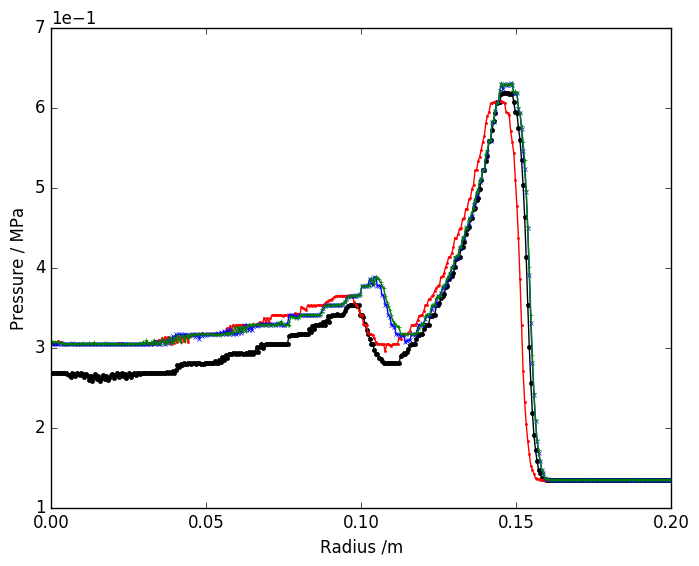} }
\caption{\label{Fig:PressureEvolution1D} Pressure evolution of a
  cylindrical plasma arc modelled with four different radiative source
  terms.  All four approaches show qualitatively the same behaviour,
  especially for the leading shock wave, which is governed primarily
  by the equation of state.  Towards the centre of the arc,
  differences between the methods are visible; the NEC approach
  initially predicts a lower pressure, whilst at late times, the grey
  body approach has a lower pressure than the three spectral methods. }
\end{figure}

\begin{figure}[!hptb]
\centering
\subfloat[\SI{20}{\micro \second} ]{\includegraphics[scale=.4]{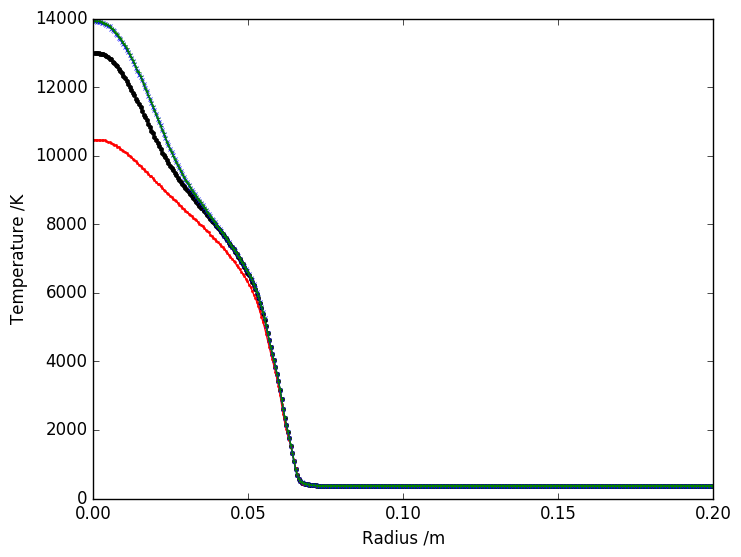} }\\

\subfloat[\SI{60}{\micro \second} ]{\includegraphics[scale=.4]{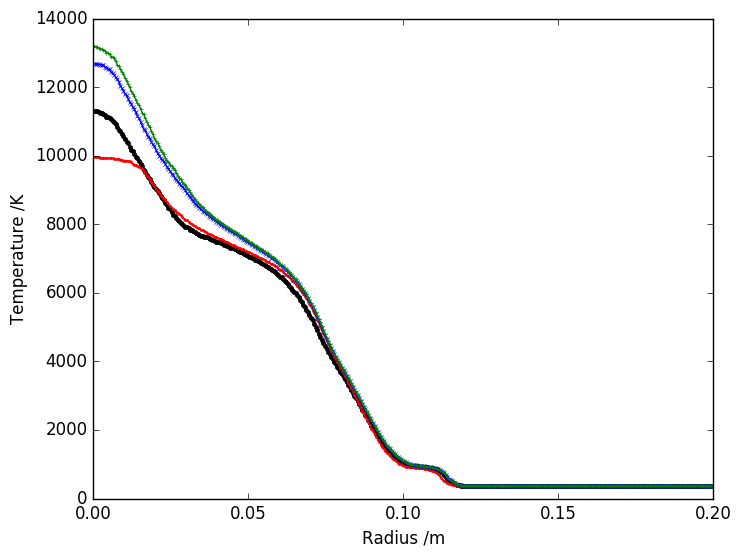} }\\

\subfloat[\SI{100}{\micro \second}]{\includegraphics[scale=.4]{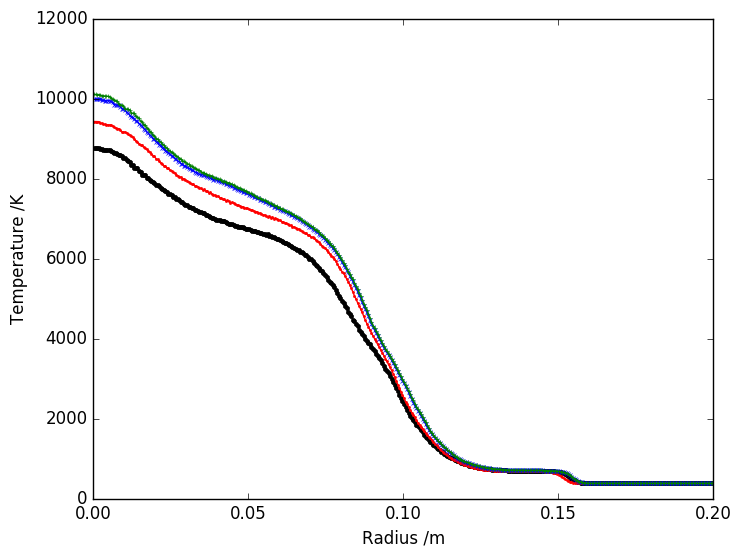} }

\caption{\label{Fig:TemperatureEvolution1D} Temperature  evolution of
  a cylindrical plasma arc modelled with four different radiative
  source terms.  The NEC approach consistently produces lower
  temperatures than the P$_1$ and SP$_3$ methods, though these
  profiles are comparable at late times.  The grey body approach is
  initially similar to the P$_1$ and SP$_3$ methods, but from around
  \SI{20}{\micro \second} onwards predicts a lower temperature.}
\end{figure}

Figures~\ref{Fig:PressureEvolution1D} and
\ref{Fig:TemperatureEvolution1D} respectively show the pressure and
temperature evolution at three output times for the grey body, NEC,
P$_1$ and SP$_3$ models.  In all cases, the pressure profiles show a
leading shock wave propagating radially outwards, followed by a strong
rarefaction.  This results in a pressure decrease at the centre of the
arc, though due to the oscillatory current input, this is not a
monotonic decrease.  Due to lower density at the centre of the arc,
the temperature remains at its highest in this region, although this
again decreases over time, as the input current reduces.

For all radiative models, the behaviour of the shock wave is largely
unchanged.  Due to the lower temperatures, the material evolution and
the equation of state dominate the behaviour here, as demonstrated by
Tr{\"a}uble et al.\cite{trauble2021improved}.  However, as is visible
in the pressure profile in figure~\ref{Fig:PressureEvolution1D}, even
at early times, after \SI{10}{\micro \second}, radiative effects are
clear at the centre of the arc.  At this time, the NEC approach
results in a lower central pressure, around \SI{0.5}{\mega \pascal}
less than the other three models.  At later times, however, the three
spectral methods result in comparable pressure profiles, whilst the
grey body approach leads to a lower pressure, 25\% lower after
\SI{150}{\micro \second}.  When comparing the three spectral methods,
the differences between the pressure profiles at early times is a
direct effect of geometric effects.  The P$_1$ and SP$_3$ take local
gradients into account in solving for the radiative source term; at
early times, when the arc radius is small, these effects are a have a
more significant contribution.  At later times, the local arc
properties dominate, and all three spectral methods have similar
behaviour.  By comparison, the simplicity of the grey body approach
shows that differences in the arc behaviour develop as a result of the
choice of radiative term.

Similar effects are visible in the temperature profile, shown in
figure~\ref{Fig:TemperatureEvolution1D}.  In this case, the grey body
approach is again comparable to the P$_1$ and SP$_3$ methods at early
times, and results in an lower temperature value at the arc centre at
later times.  Similarly, the initial low prediction of the NEC
approach is visible, though in this case, the temperature predicted by
the NEC model remains lower throughout the evolution of the arc,
though at late times it is closer to the other two spectral methods
than the grey body approach.

\section{Implementation and validation within a multiphysics
  framework}
\label{Sec:GroupCode}

This one-dimensional study suggests that the spectral methods could
offer an improvement in accuracy in modelling the centre of a
lightning arc.  To further confirm this, additional validation is
required; in particular a comparison to the behaviour at the centre of
the arc.  The results in section~\ref{Sec:NumericalResults} show that
the behaviour of the leading shock wave and the arc expansion are
largely unaffected by the choice of radiative model.  Since these are
the easiest characteristics of an arc evolution to measure
experimentally, these offer excellent validation material for the
underlying system of equations and equation of state.  In order to
extend this validation to include the radiative model, conditions at
the arc centre are required.  Such data has been recorded by
Martins~\cite{martins2016etude}, for which the thermodynamic
properties of the arc can be inferred from the optical intensity.  In
particular, Martins considers an arc attachment to an aluminium
substrate, the modelling of which requires the axisymmetric
multiphysics approach introduced in Millmore and
Nikiforakis~\cite{millmore2019multi}.

When considering the three spectral methods introduced in
section~\ref{Sec:Physics of a lightning strike}, the geometric
dependence of the P$_1$ and SP$_3$ approaches requires the solution of
multiple Helmholtz equations every time step.  In more than one
dimension, these solutions become computationally prohibitive, hence
these models are not used within this section.  However, the
performance of the NEC approach at later times suggests this may offer
improved accuracy in modelling the conditions at the centre of the
arc.  Additionally, the optically thin assumption of the NEC can be
removed by considering a decay length, as described in
section~\ref{Sec:NEC}.  In this work, a decay length of $R_p = 2$
\si{mm} was used~\cite{naghizadeh2002net}.  Since the approximate
solutions to the NEC and the reduced NEC can be tabulated based on
material properties, and included with the equation of state, these
approaches remain efficient in multiple dimensions.

\begin{figure}[!hbtp]
\centering
\includegraphics[scale=.4]{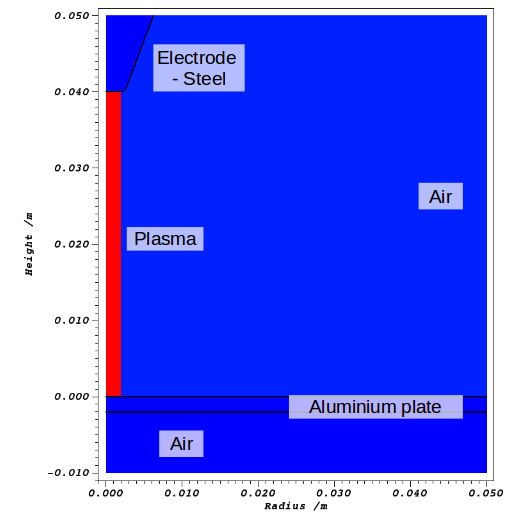}
\caption{\label{Fig:TICGroupCode} Initial geometry used in the
  multimaterial code, with the location of the pre-heated arc shown.}
\end{figure}

In order to compare to the experimental results of
Martins\cite{martins2016etude}, the configuration shown in
Figure~\ref{Fig:TICGroupCode} is used.  Within this model, there is a
geometric boundary at $r = 0$, and for the thermodynamic variables,
all other boundaries are transmissive, zero-gradient Neumann
boundaries.  For magnetic potential, the normal component to each
boundary has is a zero-gradient Neumann condition, whilst the
transverse component vanishes.  For the electric potential, all
boundaries are zero-gradient Neumann conditions except for the outer
edge of the aluminium plate, which is grounded ($\phi = 0$) and the
upper edge of the electrode, which has an input current condition.

The current input used by Martins is one of the ARP standard input
current profiles \cite{arp5412a2005aircraft}, which has an analytical
form
\begin{equation}
  I(t) = I_0 (e^{-\alpha t}- e^{-\beta t})(1 - e^{-\gamma t})^2.
  \label{Eq:ARPCurrentProfile}
\end{equation}
where the coefficients are given in table~\ref{Tab:CurrentDataMartins}.
\begin{table}[!hptb]
\begin{center}
\begin{tabular}{| c | c | c | c |}
  \hline
  $I_0$ (\si{A}) & $\alpha$ (\si{\per \second}) & $\beta$ & $\gamma$ \\ \hline
  $192039$ & $32377$ & $144719$ & $2667492$ \\
  \hline
\end{tabular}
\caption{Fitted values for the current profile described by
  equation~(\ref{Eq:ARPCurrentProfile}) as used in the experiment of
  Martins~\cite{martins2016etude}.}
\label{Tab:CurrentDataMartins}
\end{center}
\end{table}
The remaining initial conditions for these tests are given in table~\ref{Tab:ICGroupCode}.
\begin{table*}[!hptb]
\begin{center}
\begin{tabular}{| c | c | c | c |}
  \hline
  Material & $\rho$ (\si{kg m^{-3}}) & $\textbf{u}$ (\si{ms^{-1}}) & $p$ (\si{Pa}) \\ \hline
  Air ($z < -0.002$ \si{m}) &  $1.225$ & $\mathbf{0}$ & $94108$ \\ \hline
  Plate ($-0.002 < z < 0$ \si{m}) & $2710$ & $\mathbf{0}$ & $94108$ \\ \hline
  Plasma ($r < $ \SI{0.002}{m}) & $1.225$ & $\mathbf{0}$ & $4.02\times10^6$ \\ \hline
  Plasma ($r > $ \SI{0.002}{m}) & $1.225$ & $\mathbf{0}$ & $94108$ \\ \hline
\end{tabular}
\caption{Initial conditions for the attachment of a plasma arc to an
  aluminium substrate for reproducing the experiment of Martins~\cite{martins2016etude}.}
\label{Tab:ICGroupCode}
\end{center}
\end{table*}

In order to obtain thermodynamic behaviour from the optical intensity
of the plasma arc, an assumption as to the mechanism for the
generation of this emission must be made.  To achieve this,
Martins\cite{martins2016etude} develops a NEC for an arc comprising
nitrogen and oxygen, for which the singly ionised states, N$^+$ and
O$^+$, are the only radiating species.  Contributions to the
absorption coefficient are based on emission spectra for the
bound-bound contributions, but use a linear relationship for the
bound-free and free-free contributions.  Self absorption is treated
using a temperature- and pressure-dependent shell model.  Together
with the measured radiation intensity of the arc, this NEC allows arc
temperature and electron number density to be inferred; Martins states
measurement error of 5\% and 12\% respectively.  Pressure estimates
are also available, though require additional use of an equation of
state, with errors between 20\% at early times (first few
microseconds) up to 36\% at later times.

\begin{table*}[!hptb]
\begin{center}
\begin{tabular}{| c | c | c  c | c  c | c  c |}
  \hline
  & \multicolumn{7}{|c|}{Central Temperature (\si{\kilo K})} \\
  \hline
  Times (\si{ \mu s}) & Martins & Grey Body & (error) & NEC & (error)
                                                            & Reduced
                                                              NEC &
                                                                    (error) \\
  \hline
  $2$  & $37.4$ & $48.1$ & (28\%) & $18.5$ & (-50\%) & $39.9$ & (6.7\%) \\
  $4$  & $34.8$ & $54.8$ & (57\%) & $13.4$ & (-61\%) & $36.6$ & (5.1\%) \\
  $6$  & $32.8$ & $49.5$ & (50\%) & $12.5$ & (-62\%) & $30.6$ & (-6.7\%) \\
  $9$  & $28.6$ & $42.7$ & (49\%) & $12.1$ & (-58\%) & $25.5$ & (-11\%)\\
  $14$ & $25.4$ & $36.9$ & (45\%) & $12.0$ & (-53\%) & $22.4$ & (-12\%)\\
  $20$ & $22.9$ & $32.5$ & (42\%) & $11.9$ & (-48\%) & $20.9$ & (-8.7\%) \\
  $26$ & $20.7$ & $29.9$ & (44\%) & $11.8$ & (-43\%) & $21.0$ & (1.4\%)\\
  $36$ & $19.1$ & $25.4$ & (25\%) & $11.8$ & (-29\%) & $21.5$ & (9.4\%)\\
  \hline
\end{tabular}
\caption{Central arc temperatures, and errors, determined
  experimentally by Martins can be compared to those predicted by the
  grey body, NEC and reduced NEC models.  The grey body approach
  consistently over predicts temperature, whilst the NEC consistently
  under predicts temperature.  The reduced NEC shows a much closer fit
  than the other two models.}
\label{Tab:MartinsLineOutsT}
\end{center}
\end{table*}

In order to compare results from the model presented within this work,
the temperature is used, since the error measurements on this quantity
are smallest.  It is noted that the profiles shown by Martins do not
show a decay at larger radii, in contrast to other experimental
results by Hsu~\cite{hsu1983study} and other simulation results, e.g.\
Chemartin et al.~\cite{chemartin2011modelling} and Millmore and
Nikiforakis~\cite{millmore2019multi}.  The reason for this is unknown;
in order to make a consistent comparison, the central arc temperature
is used.  Results for this central temperature are shown in
table~\ref{Tab:MartinsLineOutsT}.  For the grey body approach, the arc
temperature is consistently higher than that measured experimentally,
in contrast to the results shown in
section~\ref{Sec:NumericalResults}. However, the arc itself is at a
higher temperature, and this will alter the spectral dependencies of
the radiative emission, which are not modelled under this approach,
which may account for this change in behaviour.  The NEC for an
optically thin plasma consistently under-predicts temperature, as it
did for the results shown in section~\ref{Sec:NumericalResults}.  In
this case, the higher temperatures within the arc mean that the
optically thin assumption is likely to be even less valid than for the
arc shown in figure~\ref{Fig:TemperatureEvolution1D}.  However, by
introducing an absorption length into the NEC approach, the reduced
NEC gives temperature much closer to those of Martins.  Although
errors are slightly larger than those reported by Martins, the fact
that both measurements are based on underlying physical models suggest
that this approach is performing well.  In particular, a reduced
version of the NEC approach, which can be tabulated as part of an
equation of state, can improve the accuracy of plasma arc temperatures
without excessive computational cost.  

\begin{figure*}[!hbtp]
  \centering
  \subfloat[Pressure]{\includegraphics[scale=.3]{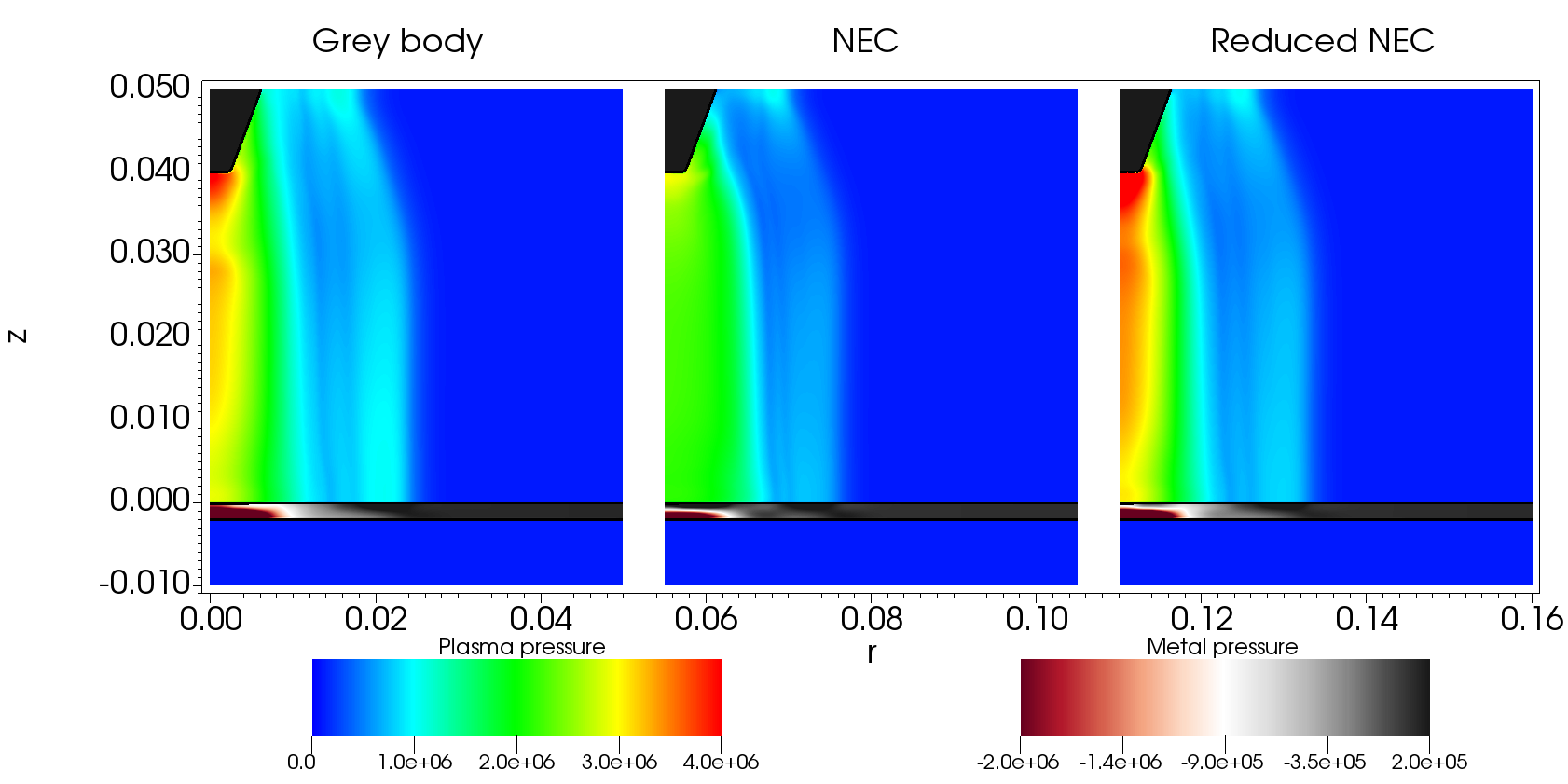}
  }  \\
  \subfloat[Temperature]{\includegraphics[scale=.3]{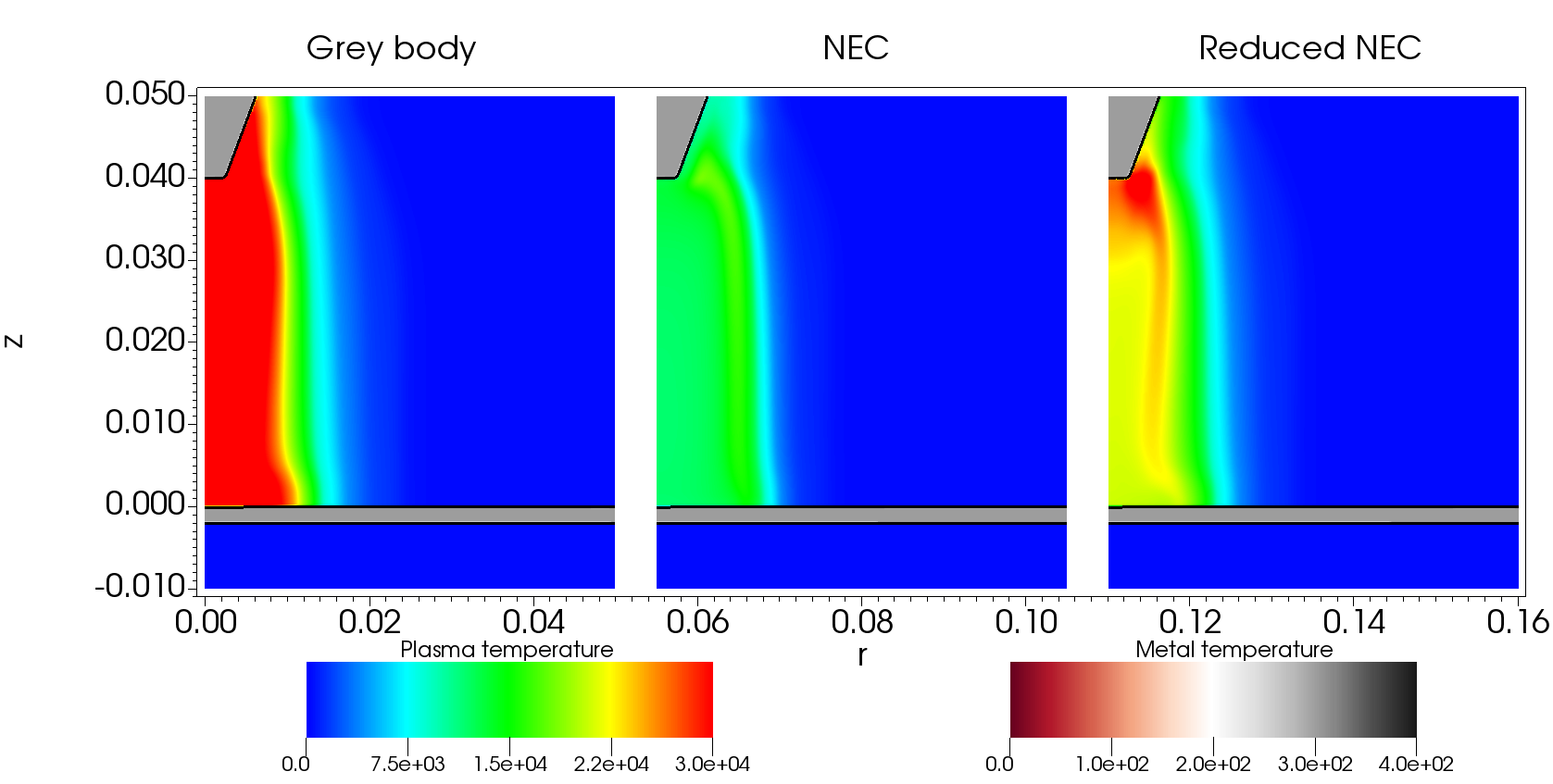} }  
  \caption{\label{Fig:TPprofiles14mus} Pressure and temperature profiles of the arc attachment to aluminium after \SI{14}{\mu s} of arc evolution showing how these vary betweeen the radiative models.  The grey body model generates the highest temperature and pressure within the arc but the reduced NEC model produces slightly higher pressures in the aluminium plate which could lead to greater deformation. }
\end{figure*}

Figure~\ref{Fig:TPprofiles14mus} shows the temperature and pressure
profiles of the arc attachment to aluminium for the three radiative
models.  In each case, the respective differences in temperature
within the arc, as described in table~\ref{Tab:MartinsLineOutsT}, are
clear within the images.  The lower temperature of the NEC approach
results in a similar lower pressure within the arc, whilst for the
reduced NEC pressure is slightly higher at the centre of the arc.  It
is believed that despite the lower central temperatures for the
reduced NEC, compared with the grey body, lower radiative emission
drives less material from the arc centre, reducing energy losses and
allowing higher pressure to be maintained. Despite these differences,
in all cases, the actual arc profile is qualitatively unchanged, as
would be expected since the bulk evolution is dependent on the
equation of state, as shown in section~\ref{Sec:NumericalResults}.

\begin{figure*}[!hbtp]
  \centering
  \subfloat[\SI{2}{\micro \second}]{\includegraphics[scale=.3]{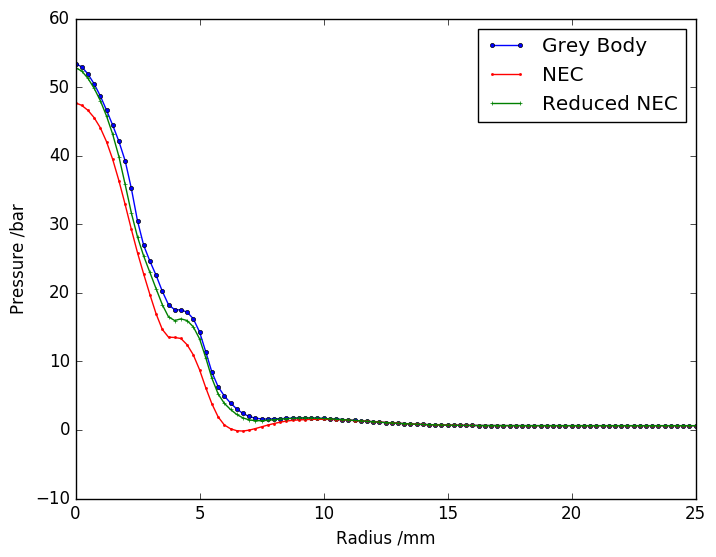} } 
  \hspace{0.5cm}
  \subfloat[\SI{4}{\micro \second}]{\includegraphics[scale=.3]{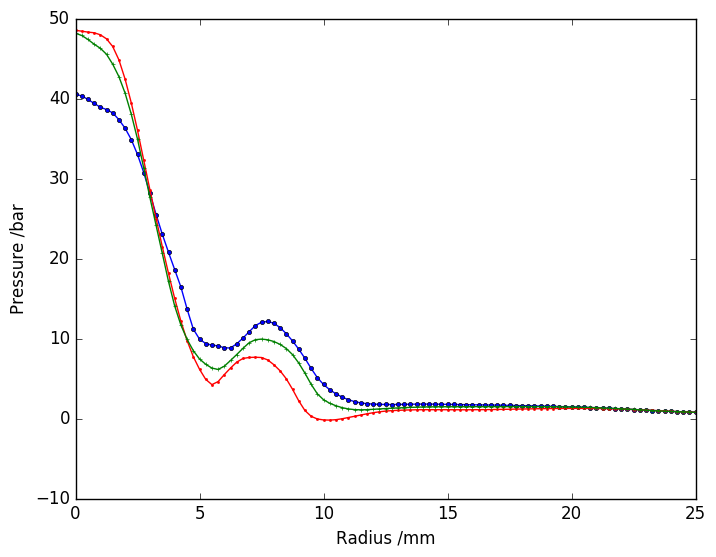} }\\
  \subfloat[\SI{6}{\micro \second}]{\includegraphics[scale=.3]{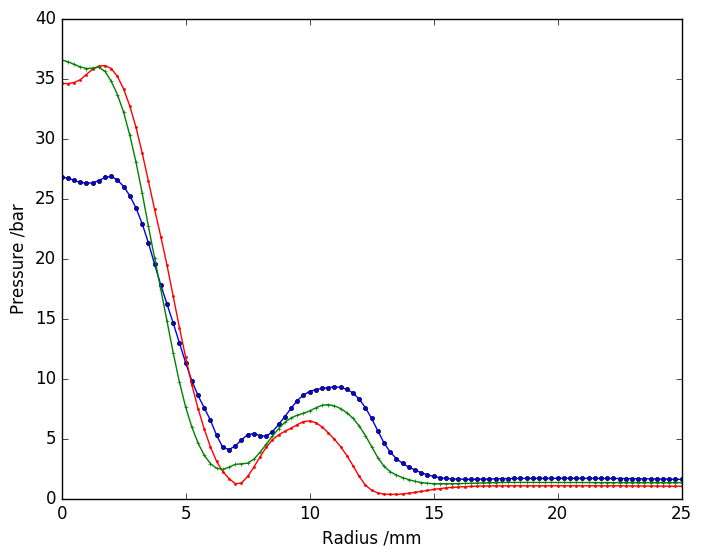} }
  \hspace{0.5cm}
  \subfloat[\SI{9}{\micro \second}]{\includegraphics[scale=.3]{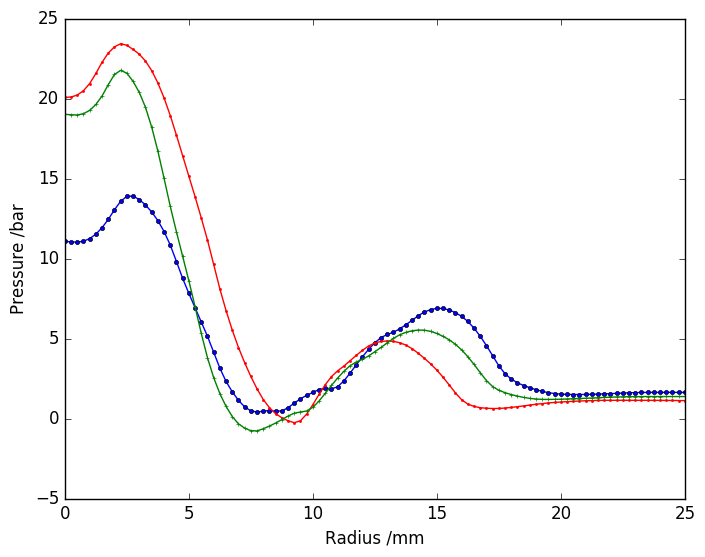} }\\
  \subfloat[\SI{14}{\micro \second}]{\includegraphics[scale=.3]{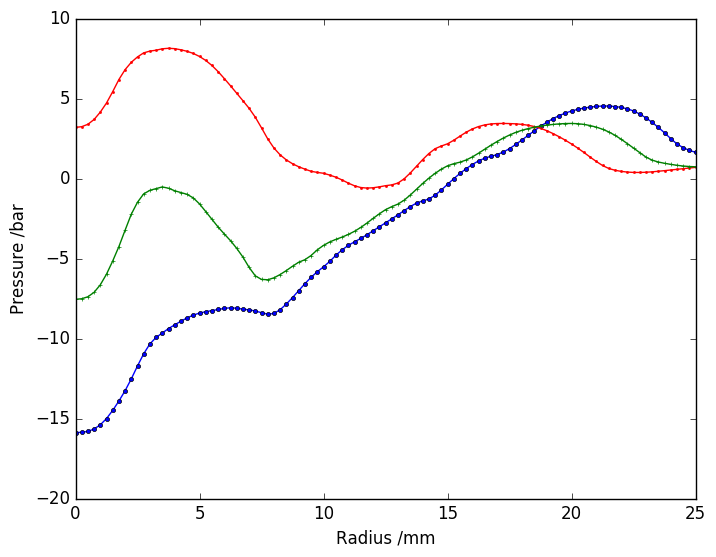} }
  \hspace{0.5cm}
  \subfloat[\SI{20}{\micro \second}]{\includegraphics[scale=.3]{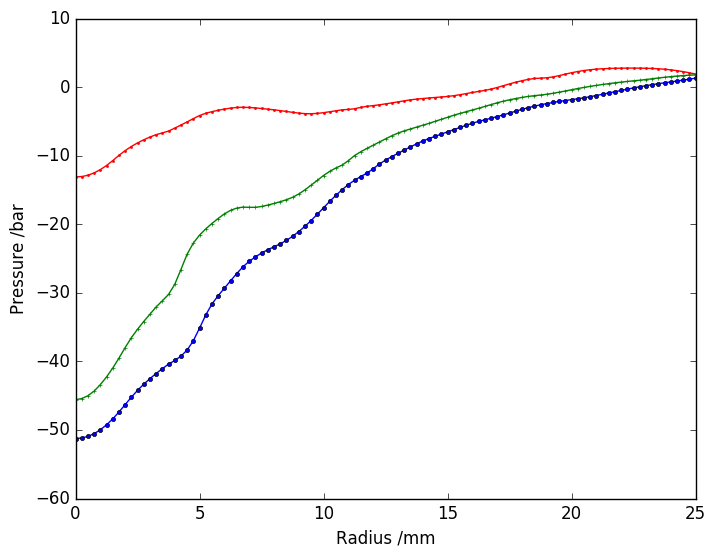} } \\
  \subfloat[\SI{26}{\micro \second}]{\includegraphics[scale=.3]{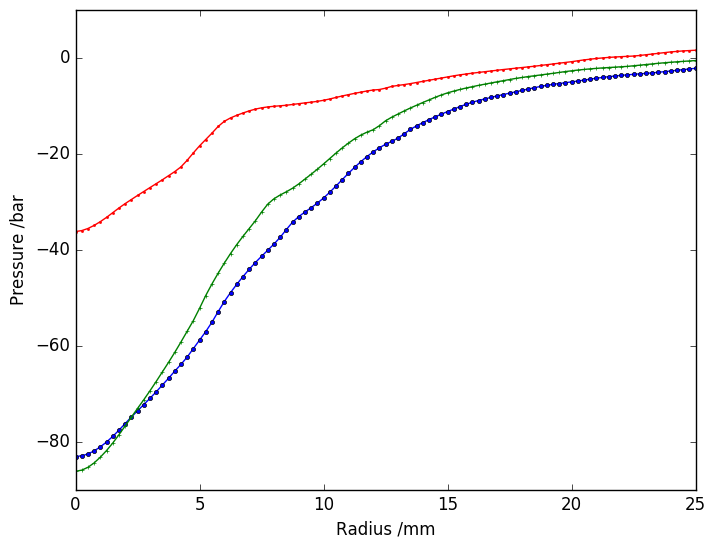} }
  \hspace{0.5cm}
  \subfloat[\SI{36}{\micro \second}]{\includegraphics[scale=.3]{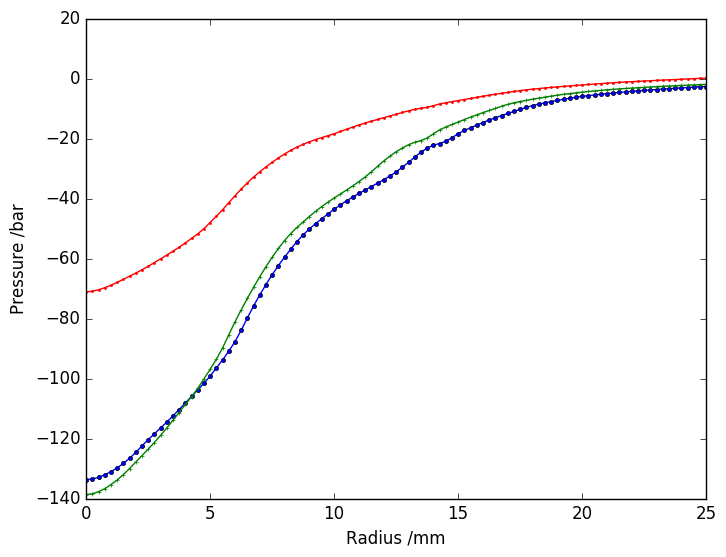} }
  \caption{\label{Fig:MartinsLineOutsPlateP} Pressure evolution at
    the surface of the aluminium plate for three radiation models
    for the attached arc.  All three show similar behaviour; an
    initial rise in pressure, followed by a drop as the substrate is
    under tension.  However, this behaviour occurs at different
    rates, with the grey body approach leading to the most rapid
    evolution of the pressure profile, and the NEC showing a much
    slower evolution.}
\end{figure*}

Although the arc behaviour is not affected by the choice of radiative
model, the attachment point to the substrate is at the centre of the
arc.  Here, the varying conditions may then have a direct effect on
the loading and damage to the substrate.  In order to investigate the
effect this may have on the substrate properties,
figure~\ref{Fig:MartinsLineOutsPlateP} shows the pressure (stress)
profile directly beneath the surface of the substrate.  For all three
models, similar qualitative behaviour is observed; an initial
high-pressure loading generates two features, an outwards moving wave,
and a subsequent, higher pressure region which remains largely
stationary.  The outwards moving wave is connected to the pressure
loading of the shock wave, whilst the conditions of the arc at the
attachment point result in a continuous loading at the centre.  At
later times, interactions within the substrate, and between the
substrate and the arc lead to a rarefaction (expansion) within the
aluminium, hence negative pressures are experienced as the material is
placed under tension.  There are clear differences between the three
radiative models, which largely coincide with the temperature
behaviour at the centre of the arc.  The outwards-moving wave moves
fastest for the grey body approach, and slowest for the NEC, and a
similar pattern is observed for the central loading.  For this latter
case, however, at late times, the grey body approach and the reduced
NEC demonstrate similar behaviour, in both cases with a higher
magnitude that the NEC approach.  It is noted that for aluminium,
these conditions, coupled with the high electrical conductivity, and
hence low Joule heating effects within the substrate, mean that
temperature does not rise more than a couple of Kelvin over this time
scale.

It is clear that the radiative model does have an effect on the
evolution of the loading on the substrate.  Quantifying this behaviour
requires a material damage model, and is beyond the scope of the
current work, which serves to highlight this possibility.  For
aluminium, it is reasonable to assume that the high electrical and
thermal conductivity of the substrate mean that these differences in
loading will have a small impact on the resultant damage behaviour.
However, for composite materials, these effects could be more
significant.  Under lightning strike, these materials can suffer
delamination, and this process starts within around \SI{10}{\micro
  \second}~\cite{KAMIYAMA201855}.  As a result, the differences in
pressure loading predicted by the radiative model may have a direct
implication for understanding material damage.

\section{Conclusions}
\label{Sec:Conclusions}

The aim of this work was to investigate the effects of the choice of
model for quantifying the radiative losses at the centre of a plasma
arc, within a multiphysics magnetohydrodynamics code.  Both the
behaviour of the arc in isolation, and when attaching to a substrate,
was considered, as was the evolution within the substrate itself.
Four radiative models were considered, all dependent on the plasma
composition, as determined by the equation of state. The simplest of
these, the grey body approach, uses uniform absorption properties and
can be computed from the bulk thermodynamic properties of the plasma,
and was initially used by Villa et al.~\cite{villa2011multiscale} The
NEC approach, which can either assume an optically thin plasma, or a
reduced approach with a finite decay length for radiative emission, is
dependent on the full plasma composition, as determined by the
equation of state, and the associated spectral absorption properties.
Two further approaches, the P$_1$ and SP$_3$ approximations, take the
geometric behaviour of the arc into account, in addition to local
properties.  These latter three approaches, referred to as spectral
methods, use a composition-dependent absorption coefficient,
calculated following the work of Chauveau et
al.~\cite{Chauveau2003RadiativeTransfer}.
  
Two key questions investigated in this work were whether the radiative
model affected the accuracy of the simulation of a plasma arc, and
whether the effects of arc attachment to a substrate were altered by
the model.  Additionally, an initial assessment into the computational
efficiency of the approaches was considered.  This considered a
one-dimensional test case, as used by Villa et
al.~\cite{villa2011multiscale} for both experimental an numerical
studies.  The results show that the three spectral methods generally
gave similar results, and these differed noticeably from the grey body
approach.  The difference between the P$_1$ and SP$_3$ was
particularly small, though at early times, the NEC approach tended to
predict lower temperatures and pressures.

In order to compare the results from the different radiative models to
experimental data, an experiment considering an arc attachment to
aluminium, conducted by Martins~\cite{martins2016etude} was
considered.  This could be approximated by two-dimensional
axisymmetry, and this introduced additional geometric complexity for
computing the P$_1$ and SP$_3$ approximations; in practice, these were
computationally inefficient.  However, the similarities between the
NEC and the P$_1$ and SP$_3$, and the fact that it could be
implemented as part of the plasma equation of state, and did not
require additional computation during a simulation, suggested that
this method could be used in the comparison to the experimental
results.  Additionally, the initial assumptions of an optically thin
plasma do not necessarily hold at the centre of a plasma arc, hence
the reduced NEC was also investigated at this point, with a finite
decay length for radiative emission.  Obtaining the properties of a
plasma arc experimentally is challenging, and requires additional
modelling for inference from optical emission, but it was shown that
the reduced NEC provided a reasonable description of the
thermodynamics of the arc centre, performing better than the grey body
and original NEC approaches.

The effects of the radiative model on the coupling between the arc and
an aluminium substrate was then considered.  This investigated the
practical application of understanding how the non-linear multiphysics
simulation of both the arc and the substrate can help understand
damage as a result of lightning strike.  Here it was shown that
pressure within the substrate over the first $\sim$\SI{30}{\micro
  \second} was dependent on the radiative model, with the grey body
approach resulting in faster pressure evolution than the spectral
approaches, and the optically thin NEC causing slower pressure
evolution.  This has potential implications for modelling damage to
low conductivity composite substrates, where it is found that
delamination of the substrate has started within these time scales.
  
This work has highlighted the importance of the choice of radiative
model for accurate simulation of lightning attachment and damage.
However, it has been shown that complex models solving for geometric
behaviour within the arc, which require the solution to multiple
Helmholtz equations within a single simulation time step, do not
provide a substantial benefit over NEC-based approaches.  Future work
based on this could improve the radiative model further based on the
NEC approach, for example improving the reduced NEC model with a decay
length based on local plasma properties.  It would also be of
significant interest to apply this approach to attachment to composite
materials, such that damage can be investigated.  This requires
additional development; composite materials are typically anisotropic,
and would require a full three-dimensional model to accomplish this,
complete with a suitable equation of state for the material, such as
that of Lukyanov~\cite{lukyanov2010equation}.  Due to the delamination
effects in this regime, this would also require a damage model to
simulate this behaviour.

In conclusion, the implementation of this equation of state-based
radiative approach offers a reliable tool to support the design and
validation of aeronautical materials under lightning strike
conditions.

\begin{acknowledgments}
  M. Apsley acknowledges support by the UK Engineering and Physical
  Sciences Research Council (EPSRC) EP/L015552/1 for the Centre for
  Doctoral Training (CDT) in Computational Methods for Materials
  Science.  The authors would also like to thank Micah Goldade of
  Boeing Research and Technology (BR\&T) for technical input
  throughout the work, and Carmen Guerra-Garcia (BR\&T and
  currently at the Massachusetts Institute of Technology) for
  suggestions on how best to model behaviours of a plasma.
\end{acknowledgments}



\appendix

\section{Absorption Coefficient}
\label{Sec:AbsorptionCoefficient}

An absorption coefficient has been developed for the 19-species EoS
used by this paper, taking into consideration electron transitions
within species of up to one degree of ionisiation, based on the
approach of Chauveau et al.~\cite{Chauveau2003RadiativeTransfer}.  The
absorption coefficient, $\kappa$, is composed of contributions from
free-free, bound-free, and bound-bound transitions
\cite{aubrecht2009net},
\begin{equation}
\kappa = \kappa^{\mathrm{bb}} + \kappa^{\mathrm{bf}} + \kappa^{\mathrm{ff}}.
\end{equation}

The bound-bound contributions produce spectral lines when electrons
move between bound states in an atom or molecule, where the atomic
lines depend on wavelength, $\lambda$, and temperature, $T$, accroding to~\cite{martins2016etude}
\begin{multline}
  \kappa^{bb} = \frac{\lambda^5(e^{hc/\lambda k_B T} - 1)}{2hc^2} 
  \sum_{\mathrm{line}} \frac{hc}{4\pi} \left( \frac{g_u A_{ul}}{\lambda_{\mathrm{line}}} \right) \\
   \times \frac{N_0(T,N_e)}{Q(T)} e^{-E_u/k_B T} \phi (\lambda- \lambda_{\mathrm{line}}, T, N_e)
 \end{multline}
 where $e$ is the charge on an electron, $h$ Planck's constant, $c$
 the speed of light in a vacuum, $k_B$ the Boltzmann constant, $g_u$
 ane $E_u$ the degeneracy and energy of the upper energy level,
 $A_{ul}$ the Einstein absorption coefficient for the transition,
 $N_0$ and $N_E$ the number densities of the radiating species and the
 electrons respectively and
 $\phi(\lambda- \lambda_{\mathrm{line}},T,N_e)$, is the spectral line
 shape.  A spectral line always has a natural line width because there
 is a quantum-related uncertainty in the energy of the states
 involved, though for the scales considered in this work, the natural
 line width of a plasma, around $\sim$\SI{e-4}{\nano \metre},can be
 neglected.  For an air plasma, the Stark effect due to collisions
 between species is the dominant line-broadening effect.  Doppler
 shifts due to thermal motions of the atoms can also cause line
 broadening in the shape of a Gaussian distribution of half width at
 half maximum (HWHM) of $\sim$ \SI{0.01}{\nano \metre} at temperatures
 of \SI{40 000}{K}.  This is approximately $20$ times smaller than the
 HWHM of the Stark contribution, thus Doppler effects can be
 neglected. Also negligible are resonance and Van der Waals
 contributions which have HWHM proportional to the density of neutral
 perturbers which decreases sharply with temperature
 \cite{aubrecht2009net}.

 The line shape of the Stark effect for a line at wavelength
 $\lambda_{\mathrm{line}}$ and HWHM of $\gamma$ is approximated by a
 Lorentzian profile,
\begin{equation}
  \phi(\lambda - \lambda_{line}, T, N_e) = \frac{1}{\pi} \frac{\gamma}{\gamma^2 + (\lambda - \lambda_{\mathrm{line}})^2}.
\end{equation}
In this work, four species were considered; O, N, O$^+$, and N$^+$.
The data for O$^+$ and N$^+$ was taken from appendix B of
Martins~\cite{martins2016etude}, the line widths for N and O are
tabulated in the work of Konjecvi{\'c} and
co-authors~\cite{konjevic2002experimental,konjevic1976critical,konjevic1984experimental},
and the rest of the required data was taken from the NIST database
\cite{NIST_ASD}.  The Stark line width was assumed to be proportional
to number density of electrons and inversely proportional to the
square root of temperature.

Bound-bound contributions for diatomic molecules take the form~\cite{owen2014measurements}
\begin{equation}
  \kappa^{\mathrm{bb}}_{\mathrm{mol}} = N_i L \sum_j S_{ij} \phi_{ij}
\end{equation}
where $N_i L$ is the number density of energy level $i$ for each
molecule.  The line strength $S_{ij}$ of the transition from $i$ to
$j$ is \cite{gordon2017hitran2016}
\begin{equation}
\begin{aligned}
  S_{ij} &= \frac{A_{ij}}{8 \pi c \nu_{ij}^2} g' \frac{e^{-c_2E''/T}(1-e^{-c_2 \nu_{ij}})}{Q(T)} \\
  	&= S_{ij}(T_{\mathrm{ref}}) \frac{Q(T_{\mathrm{ref}})}{Q(T)} \frac{\exp(-c_2E''/T)}{\exp(-c_2E''/T_{\mathrm{ref}})} \\
  	&\times \frac{(1 - \exp(-c_2 \nu_{ij}/T))}{(1 - \exp(-c_2 \nu_{ij}/T_{\mathrm{ref}}))}
\end{aligned}
\end{equation}
where $c_2 = hc/k_B$, $Q(T)$ is the partition function, $E''$ is the
energy of the lower level, $\nu_{ij}$ is the frequency of the
transition and $A_{ij}$ is the Einstein emission coefficient.  This is
only strictly valid for temperatures near $T_{\mathrm{ref}} =$
\SI{296}{\kelvin} however results obtained for temperatures up to
\SI{15 000}{\kelvin} are sufficiently accurate for the purposes of
this study and molecular bound-bound contributions are only
significant for $T <$ \SI{10 000}{\kelvin}.  Chauveau et
al.~\cite{Chauveau2003RadiativeTransfer} show that for the NEC only
N$_2$, NO and O$_2$ contribute. The data for N$_2$ and O$_2$ were
obtained from the HITRAN molecular spectral database
\cite{gordon2017hitran2016}.  No data for NO could be found.
  
Next the free-free contributions were considered. For N$^+$ and O$^+$
the emission coefficient for a plasma with ion density $N_i$ and
electron density $N_e$ is
\begin{equation}
  \varepsilon_{\nu}^{\mathrm{ff}} = g(\tilde{\nu},T) \frac{8}{3} \left( \frac{2 \pi}{3 k_B T m_e} \right)^{1/2} \frac{\alpha^2 e^6}{m_e c^3} \exp \left( - \frac{h \nu}{k_B T}  \right) N_e N_i.
\end{equation}
where $\alpha$ is the fine structure constant and $m_e$ is the electron mass.  For $T >$ \SI{11604}{\kelvin} the expression for the Gaunt factor, $g(\tilde{\nu},T)$ \cite{stallcoP1974analytical},
\begin{equation}
  \begin{aligned}
  g(\lambda,T) &= g_c(\lambda,T)[1+\varepsilon_c(\lambda,T)] \\
  g_c(\lambda,T) &= 1.270 \ln (0.2066 \lambda T^{3/2}) \\
  \varepsilon_c &= - \left(7.4 - \frac{45}{T}\right) \frac{1}{\lambda} + \left(4.9 - \frac{18}{T} \right) \frac{100}{\lambda^2}
  \end{aligned}
\end{equation}
was assumed.  For $T <$ \SI{11604}{\kelvin} the Gaunt factor of Menzel
and Pekeris~\cite{menzel1935absorption} was used: 
\begin{equation}
  g_{\nu}(\nu,T) = 1 + 0.1728 \left( \frac{\nu}{R_H Z^2} \right)^{1/3} \left[ \frac{2}{\kappa^2} \left( \frac{R_H Z^2}{\nu}\right) - 1 \right],
\end{equation}
where $Z$ is the atomic number and $R_H$ is the Rydberg constant.

For N and O, absorption occurs through inverse bremsstrahlung whereby
free electrons absorb radiation.  The absorption coefficient is then
given by
\begin{multline}
  \kappa^{\mathrm{bb}} = 8 \pi^2 (2\pi)^{1/2} \left( \frac{\hbar c}{e^2} \right)^2 \frac{(e^2/a_0)^{3/2}(k_BT)^{3/2}}{(h\nu)^3} \\
    \times \left[ 1 + \frac{h\nu}{2k_B T}\right] a_0^5 N N_e
\end{multline}
where $a_0$ is Bohr's constant and $\hbar = \frac{h}{2\pi}$.  This is
valid for wavenumbers in the range \numrange{4800}{24 000}\si{\per
  \centi \metre} and temperatures of \numrange{4000}{12
  000}\si{\kelvin}.

The molecular contributions to the free-free component also arise from inverse bremsstrahlung and depend on the cross section $\Sigma_a(\tilde{\nu},T)$ according to
\begin{equation}
  \kappa^{\mathrm{ff}} = \Sigma_a(\tilde{\nu},T) N_{mol} \left[ 1 - \exp \left( - \frac{hc}{\lambda k_B T} \right) \right]
\end{equation}
The contribution from molecular ions is negligible
\cite{Chauveau2003RadiativeTransfer} and so only N$_2$ and O$_2$ are
taken into account.  The data from Kivel~\cite{kivel1967bremsstrahlung}, which is given for
wavelengths of \numrange{0.3}{4.8} \si{\micro \metre} and temperatures
of \numrange{3 000}{15 000} \si{\kelvin}, was interpolated within this
range and extrapolated for $T <$ \SI{3000}{\kelvin} or wavenumbers,
$\tilde{\nu} <$ \SI{2083}{\per \centi \metre}.

The bound-free continuum absorption coefficient also has atomic and
molecular contributions.  For N, O, N$^+$ and O$^+$ it has the form
\begin{equation}
\label{Eq:BFContAbsCoeff}
  \kappa^{\mathrm{bf}} = \left[ \sum_i N_i \Sigma_i(\lambda) \right]\left[ 1 - \exp \left( - \frac{hc\tilde{\nu}}{k_B T} \right) \right]
\end{equation}
where $g_i$ is the statistical weight of level $i$, and the population density of level $i$ depends on the total population density $N_0$ according to
\begin{equation}
  N_i = N_0 \frac{g_i}{Q(T)} \exp \left( -\frac{\Delta E_i}{k_B T} \right)
\end{equation}
and the partition function is
\begin{equation}
  Q(T) = \sum_i g_i \exp \left( -\frac{\Delta E_i}{k_B T} \right)	.
\end{equation}  
The transition energies, $\Delta E$, and cross sections, $\Sigma_i$, are taken from the TOPBASE database \cite{TOPBASE}. 

For O$^-$ and N$^-$ photoabsorption by the $^3P$ and $^2P$ levels
respectively dominates with population densities also given by
equation~\eqref{Eq:BFContAbsCoeff}.  The energy levels for N$^-$ can
be found in Cowan et al.~\cite{cowan1997excited} and for O$^-$ they
are given by Garrett and Jackson Jr.~\cite{garrett1967electron}. The
absorptivity is then
\begin{equation}
  \kappa_{\sigma} = N \Sigma(\lambda) \left[ 1 - \exp \left( -\frac{hc}{\lambda k_B T} \right) \right]
\end{equation}
where for N$^-$
\begin{equation}
  \Sigma = \begin{cases} 2 \times 10^{-17} & \text{if  } \tilde{\nu} \geq 806.55 \text{cm}^{-1} \\
  			0 & \text{Otherwise} \end{cases}
\end{equation}
and for O$^-$ the values of $\Sigma$ come from \cite{Chauveau2003RadiativeTransfer}.

The molecular contributions, from N$_2$, O$_2$ and NO, have the same
form as for photoabsorption.  In the case of N$_2$ and O$_2$ the cross
sections are taken from Fennelly and
  Torr~\cite{fennelly1992photoionization}.  For NO, an analytical
expression can be found in Romanov et
  al.~\cite{romanov1995thermodynamic}. The data determining the
shape of the Schumann-Runge continuum, the longest absorption band in
molecular oxygen, are taken from  Churchill et
  al.~\cite{churchill1966absorption} for temperatures in the range
\numrange{300}{1000}\si{\kelvin} and wavenumbers of \numrange{57
  000}{77 000} \si{\per \centi \metre}.  This is extrapolated linearly
down to $\tilde{\nu} =$ \SI{48 000}{\per \centi \metre}.

\begin{figure*}[!hbtp]
\center
  \includegraphics[scale=.65]{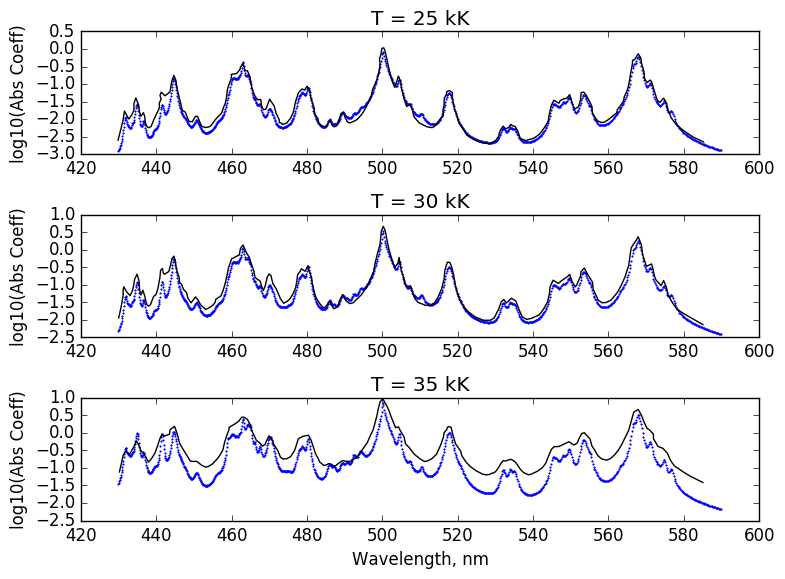} 
  \caption{\label{Fig:VariationWithT} Bound-bound absorption
    coefficient (cm$^{-1}$) for N$^+$ and O$^+$ for
    $N_e =$\SI{3e18}{\per \cubic \centi \metre} for temperatures of
    \SI{25}{\kilo \kelvin}, \SI{30}{\kilo \kelvin}, and \SI{35}{\kilo
      \kelvin}; calculated results in blue, results from
    Martins~\cite{martins2016etude} in black.  These show good
    agreement, especially for lower arc temepratures, though some
    differences in absorption are visible for higher temperatures.}
\end{figure*}

The atomic bound-bound contributions were validated against results
from Martins \cite{martins2016etude} in
figure\ref{Fig:VariationWithT}.  Sample results are shown for $N_e =$
\SI{3e18}{\per \cubic \centi \metre} at temperatures of \SI{25}{\kilo
  \kelvin}, \SI{30}{\kilo \kelvin}, and \SI{35}{\kilo \kelvin}.  The
results match well, except at \SI{35}{\kilo \kelvin} where the peaks
coincide but the troughs are deeper.  This may be due to the
difference in number density between the estimated value and the true
value used by Martins.

\begin{figure*}[!hbtp]
\centering
\subfloat[]{\includegraphics[scale=.35]{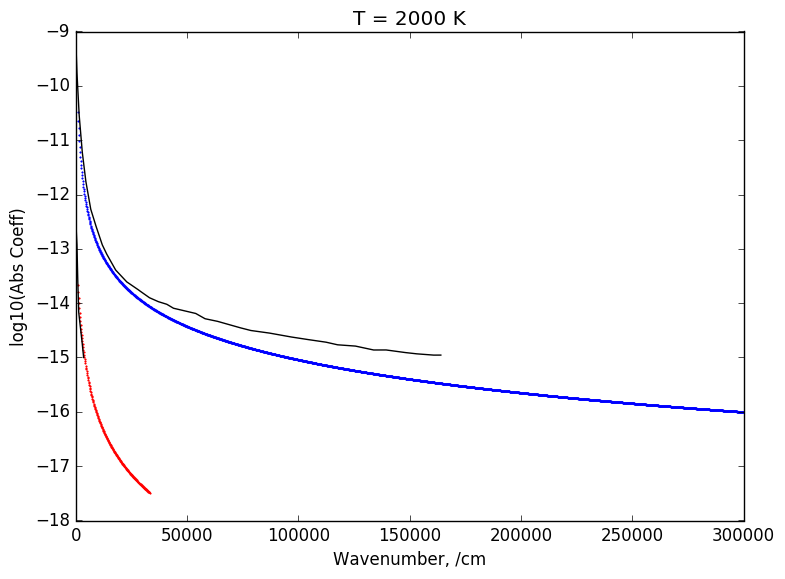} }
\subfloat[]{\includegraphics[scale=.35]{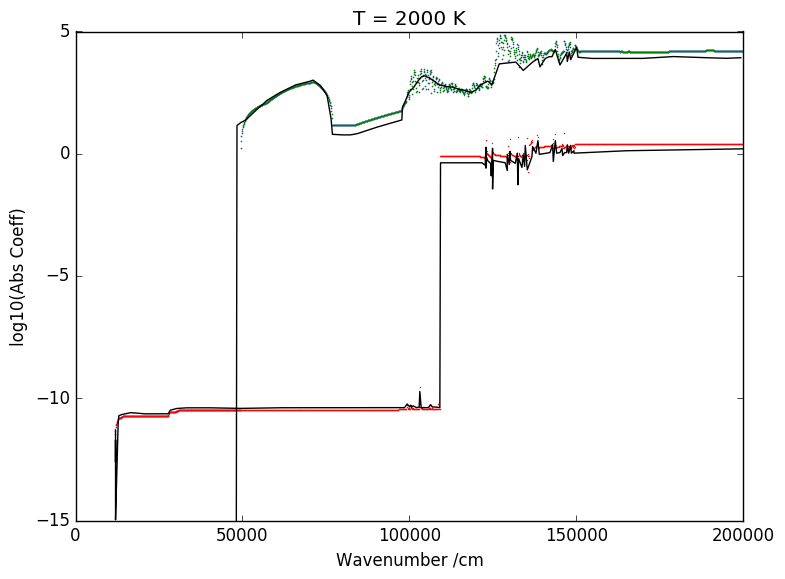} } \\
\subfloat[]{\includegraphics[scale=.35]{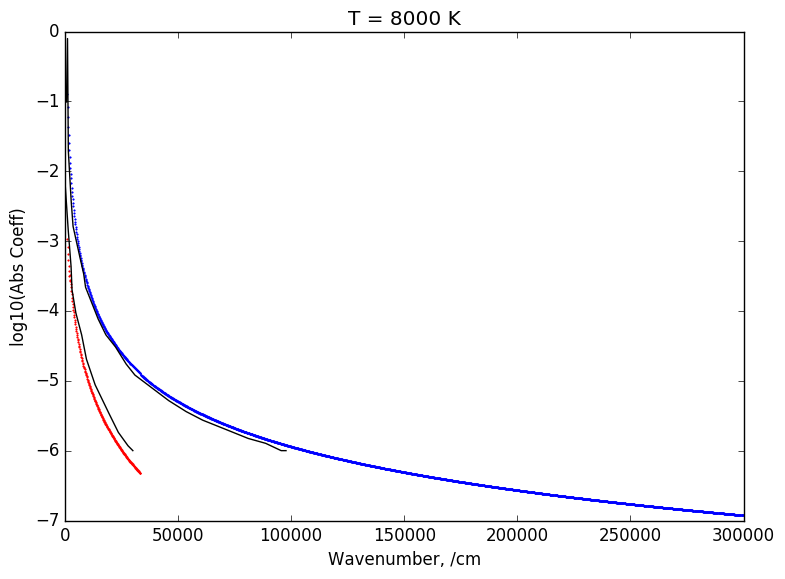} }
\subfloat[]{\includegraphics[scale=.35]{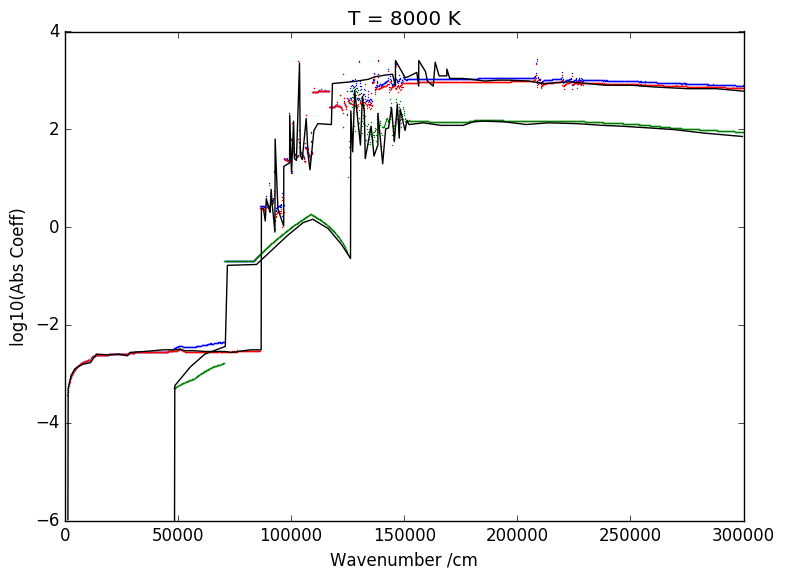} } \\
\subfloat[]{\includegraphics[scale=.35]{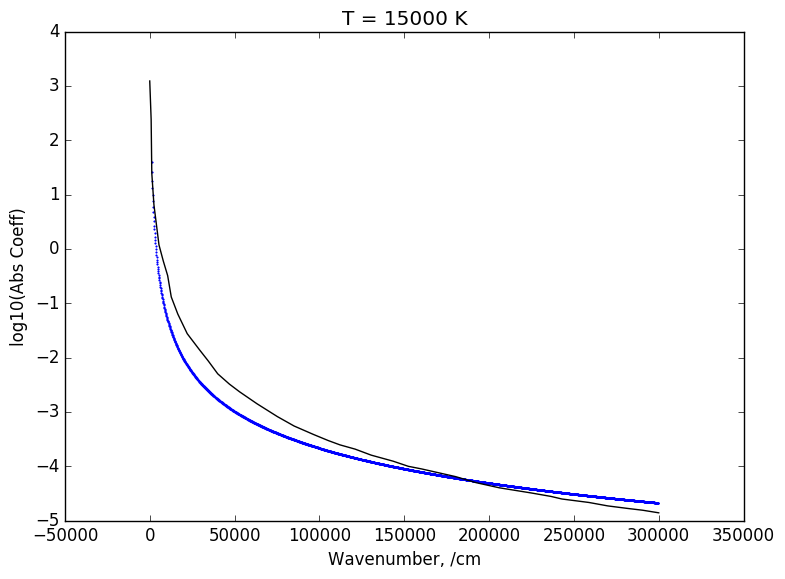} }
\subfloat[]{\includegraphics[scale=.35]{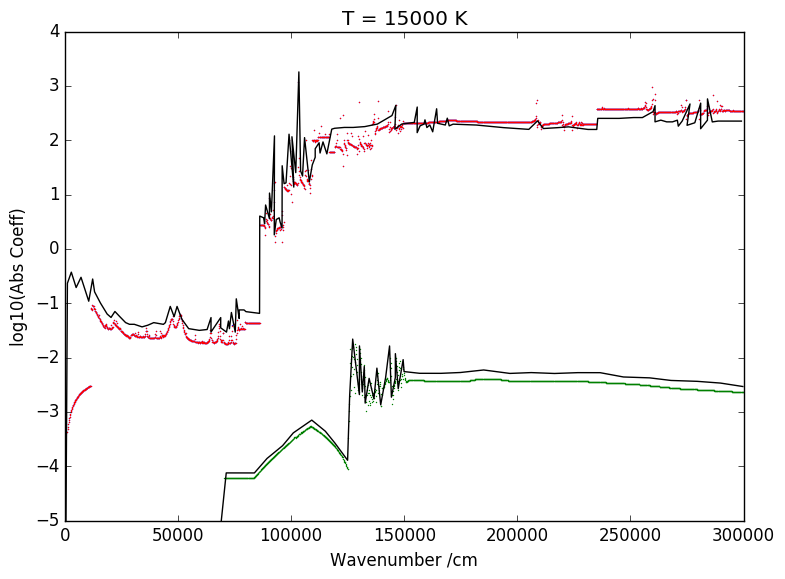} }  
\caption{\label{Fig:ContinuumAbsCoeff} Free-free (left) and bound-free
  (right) contributions (molecular in red, atomic in blue) compared to
  results of Chauveau et al.~\cite{Chauveau2003RadiativeTransfer} at
  (top) \SI{2000}{\kelvin}, (middle) \SI{8000}{\kelvin}, (bottom)
  \SI{15 000}{\kelvin}.  Comparable values for the absorption
  coefficinet are visible, though at higher temperatures, some peaks
  are missing.}
\end{figure*}

To validate the continuum absorption coefficients, the bound-free and
bound-bound contributions are plotted in figure
\ref{Fig:ContinuumAbsCoeff} for temperatures of
\numlist{2000;8000;15000}\si{\kelvin} against the results obtained by
Chauveau et al.~\cite{Chauveau2003RadiativeTransfer}.
Molar densities from the NASA CEA database \cite{mcbridenasa} are
used. They agree well and only diverge significantly at wavelengths
where the contributions are no longer significant.

\begin{figure}[!hbtp]
\centering
\subfloat[]{\includegraphics[scale=.33]{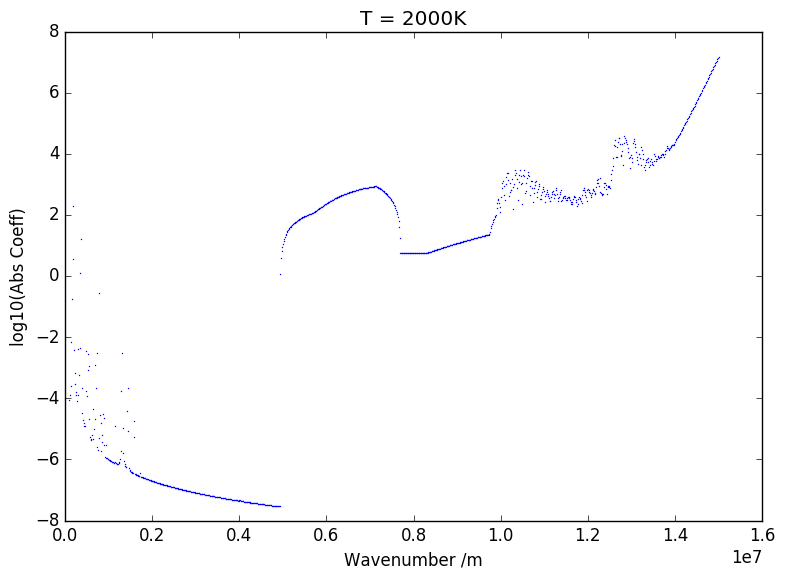} }

\subfloat[]{\includegraphics[scale=.33]{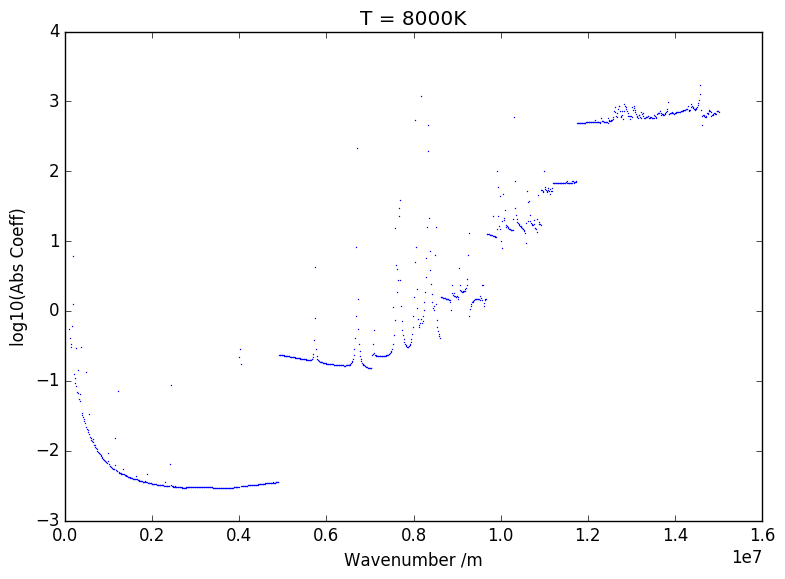} }

\subfloat[]{\includegraphics[scale=.33]{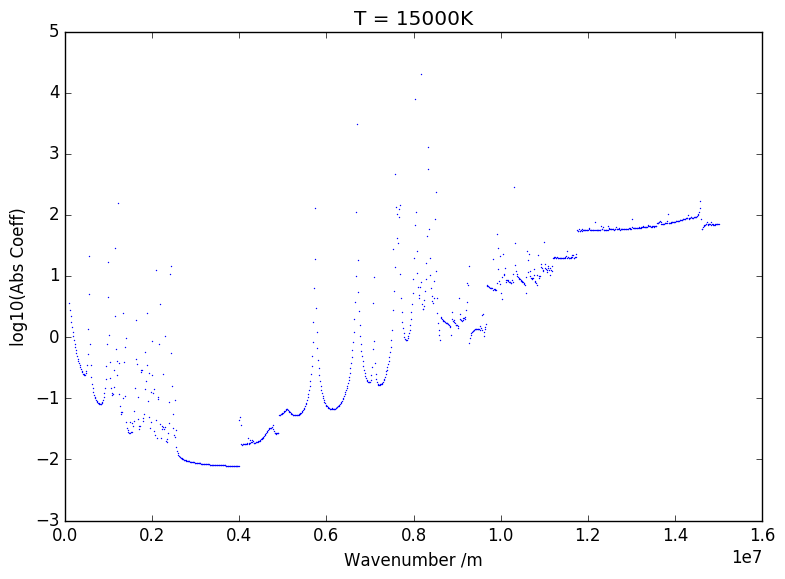} }

\caption{\label{Fig:TotalKappa} Total absorption coefficient (left) at
  1 atm (top) \SI{2000}{\kelvin}, (middle) \SI{8000}{\kelvin},
  (bottom) \SI{15 000}{\kelvin} in all three cases, these results
  compare well with those of Chauveau et
  al.~\cite{Chauveau2003RadiativeTransfer}.  Although some structures,
  including the Schumann-Runge bands, are not present in these
  results, the comparable shape and peak values suggest that this is a
  reasonable model for absorption coefficient.}
\end{figure}

Figure \ref{Fig:TotalKappa} shows the total absorption coefficient.
The results compare well to those of Chauveau et
al.~\cite{chauveau2002contributions} though there are some differences
due to the fact that the HITRAN database does not include data for as
many lines as are considered by Chauveau et al.\ and consequently some
of the Schumann-Runge bands are missing at \SI{2000}{\kelvin}.

\bibliography{ms}

\end{document}